\documentclass[twocolumn]{svjour3}       
\usepackage{booktabs} 

\usepackage{color}
\usepackage{algpseudocode}
\usepackage{algorithm}

\usepackage{amsthm}
\usepackage{amsmath}
\usepackage{cancel}
\usepackage{graphicx}
\usepackage{hyperref}
\usepackage{tabularx}
\usepackage{ragged2e}  
\usepackage{textcomp}
\usepackage{url}
\usepackage[labelfont=bf]{caption}
\usepackage[caption=false]{subfig}
\usepackage[table,xcdraw]{xcolor}

\newcolumntype{Y}{>{\RaggedRight\arraybackslash}X}

\DeclareMathOperator*{\argmax}{arg\,max}

\newcommand{\squishlist}{
 \begin{list}{$\bullet$}
  { \setlength{\itemsep}{0pt}
     \setlength{\parsep}{1pt}
     \setlength{\topsep}{1pt}
     \setlength{\partopsep}{0pt}
     \setlength{\leftmargin}{1em}
     \setlength{\labelwidth}{1em}
     \setlength{\labelsep}{0.5em} } }

\newcommand{\squishend}{
  \end{list}
}

\definecolor{americanrose}{rgb}{1.0, 0.01, 0.24}
\definecolor{airforceblue}{rgb}{0.36, 0.54, 0.66}
\definecolor{ao(english)}{rgb}{0.0, 0.5, 0.0}
\definecolor{ao}{rgb}{0.0, 0.0, 1.0}

\newcommand{\rjmX}[1]{\textcolor{black}{#1}}

\newcommand{\ch}{\operatorname{ch}}
\newcommand{\parent}{\operatorname{par}}
\newcommand{\eat}[1]{}

\newcommand{\dom}{\mathrm{dom}}
\newcommand{\data}{\mathrm{data}}

\newcommand{\attD}{\ensuremath{T}}
\newcommand{\attA}{\ensuremath{A}}

\newcommand{\child}{\mbox{$c$}}

\begin{document}

\title{Data Lake Organization}

\author{Fatemeh Nargesian \and
Ken Q. Pu \and
Bahar Ghadiri Bashardoost \and
Erkang Zhu \and
Ren\'ee  J. Miller
}

\institute{Fatemeh Nargesian \at
        University of Rochester \\
	\email{f.nargesian@rochester.edu} \and 
        Ken Q. Pu \at
        University of Ontario Institute of Technology \\
	\email{ken.pu@uoit.ca} \and 
	Bahar Ghadiri Bashardoost \at
        University of Toronto \\
	\email{ghadiri@cs.toronto.edu}  \and 
	Erkang Zhu \at
        Microsoft Research \\
	\email{ekzhu@microsoft.com} \and 
	Ren\'ee  J. Miller \at
	Northeastern University \\
	\email{miller@northeastern.edu}  
}

\maketitle

\begin{abstract}
We consider the problem 
of creating a  navigation structure that allows a user to most
effectively navigate a data lake. 
We define an {\em organization} as a graph that contains 
nodes representing sets of attributes within a data lake and edges indicating subset 
relationships among nodes.  
We present a new probabilistic model of how
users interact with an organization 
and define the likelihood of a user finding a table using the organization. 
We propose the {\em data lake organization problem} as the problem 
of finding an organization that maximizes the expected probability of discovering tables 
by navigating an organization. 
We propose an approximate algorithm for the data lake organization problem. 
We show the effectiveness of the algorithm on both real data lakes
	containing data from \eat{federal} open data portals and on benchmarks that 
emulate the observed characteristics of real data lakes. 
	Through a formal user study, 		we show that 
		navigation can help users discover  
		relevant tables that cannot be found by keyword search. 
In addition, in our study, 
42\% of users preferred the use of navigation and 58\% preferred keyword search, 
suggesting these are complementary and both useful 
		modalities for data discovery in data lakes. 
Our experiments show that data lake organizations 
take into account the data lake distribution 
	and outperform an existing hand-curated taxonomy 
	and a common baseline organization. 
\end{abstract}

\section{Introduction}
\label{sec:introduction}

The popularity and growth of data lakes 
is fueling interest in 
dataset
discovery.
Dataset discovery is normally formulated as a search problem.  
In one version of the problem, the query is a set of keywords and the goal is to 
find tables relevant to the keywords~\cite{DBLP:conf/www/BrickleyBN19,Pimplikar:2012}. 
Alternatively, the query can be a table (a query table), 
and the problem is to find other tables that are close
to the query table~\cite{Cafarella:2009}. 
If the input is a query table, 
then the output may be tables  that join 
or union with query table~\cite{NargesianZPM18,DasSarma:2012,Zhu:2016,ZhuDNM19,YangZYZLY18}.

A complementary alternative to search is
navigation.  In this paradigm, a user navigates through 
an organizational
structure to find tables of interest.  
In the early days of Web search, navigation was the dominant discovery
method for Web pages.  
Yahoo!, a mostly hand-curated directory structure, was the most significant
internet gateway for Web page discovery~\cite{Yahoo}. 
\footnote{It is interesting to note that Yahoo! may stand for 
``Yet Another Hierarchical Officious (or Organized) Oracle''.}
Even today, 
hierarchical organizations of Web content (especially
entities like videos or products) is still 
used by companies such as \texttt{Youtube.com} and
\texttt{Amazon.com}.
Hierarchical navigation allows  
a user to browse available 
entities going from more general concepts to
more specific concepts using existing 
ontologies or structures automatically created using
taxonomy induction~\cite{Snow:2006:STI,Kozareva:2010,Yang:2009}. 
When entities have known features, we can apply faceted-search over entities~\cite{zheng2013survey,KorenZL08,velardi2013ontolearn}.
Taxonomy induction looks 
for {\em is-a} relationships between entities (e.g., {\tt student is-a person}), 
faceted-search applies 
predicates (e.g., {\tt model = volvo}) to filter entity collections~\cite{Arenas:2016:FSO}.  
In contrast to hierarchies over entities, 
in data lakes tables contain attributes that may mention many different types of entities and relationships between them. 
There may be {\em is-a} relationships between tables, or their attributes, and no easily defined facets for grouping tables. 
If tables are annotated with class labels of a knowledge base (perhaps using entity-mentions in attribute values), 
the {\em is-a} relationships between class labels could provide an organization on tables.
However, recent studies show the difference in size and coverage of domains in various public cross-domain 
knowledge bases and highlight that no single public knowledge base sufficiently covers all
domains  and attribute values represented in heterogeneous corpuses such 
as data lakes~\cite{hassanzadeh2015understanding,Ritze:2016:PPW:2872427.2883017}. 
In addition, the coverage of standard knowledge bases on attribute values in open data lakes is extremely low~\cite{NargesianZPM18}. 
Knowledge bases are also not designed to provide effective navigation. 
We propose instead to build an organization that is designed to
  best support navigation and exploration over heterogeneous data lakes. 
Our goal is not to compete with or replace search, but rather to
  provide an alternative discovery option for users with only a vague
  notion of what data exists in a lake. 

\subsection{Organizations}
\label{sec:orgs}

Dataset or table search is often done using attributes by finding similar, joinable, 
or unionable attributes~\cite{DasSarma:2012,fernandez2018aurum,fernandez2018seeping,NargesianZPM18,Zhu:2016,ZhuDNM19}. 
We follow a similar approach and 
define an {\em organization} as 
a Directed Acyclic Graph (DAG) 
with nodes that represent sets of attributes
from a data lake.  A node may have a label created from the attribute values
  themselves or from metadata when available.  
A table can be associated to all nodes containing one or more of its attributes.
An edge in this DAG indicates that the attributes in a parent node is a superset of the
attributes in a child node. 
A user finds a table by traversing a path from 
a root of an organization 
to any leaf node that contains any of its attributes.  

We propose the {\em data lake organization problem}
where the goal is to find an organization that allows a user to most
efficiently find tables.  
We describe user navigation of an organization using a Markov model. 
In this model, each node in an organization is equivalent to a state in the navigation model. 
At each state, a user is provided with a set of next states (the children of the current state).
An edge in an organization indicates the transition from the current state to a next state.  
Due to the subset property of edges, each transition filters out some attributes until the navigation reaches
attributes of interest. 
An organization is effective  if certain properties hold. 
At each step, a user should be presented with 
only a reasonable number of choices.
We call the maximum number of choices the {\em branching factor}.
The choices should be distinct (as dissimilar as possible) to make it easier for a user to choose 
the most relevant one. 
The transition probability function of our model 
assumes users choose the next state that has the highest similarity to the topic query they have in mind. 
Also, the number of choices they need to make (the {\em length of the discovery path}) 
should not be large. 
Furthermore, in real data lakes, as we have observed and report in detail in our evaluation 
(Section~\ref{sec:evaluation}), the topic distribution is typically skewed (with a few tables on some topics and 
a large number on others). 
Hence, give the typical skew of topics in real data lakes, 
a data lake organization must be able to automatically determine over which portions of the data, 
more organizational structure is required, and where a shallow structure is sufficient.

\begin{table}[h]
\small
\centering
	\caption{Some Tables from Open Data.} 
\label{tbl:examplerepo}
	\begin{tabular*}{\columnwidth}{@{}  l  c  @{} }
\toprule
	\textbf{Id} & \textbf{Table Name} \\
\midrule
	 d1 & {\em Surveys Data for Olympia Oysters, Ostrea lurida, in BC}\\
	 d2 & {\em Sustainability Survey for Fisheries}\\
	 d3 & {\em Grain deliveries at prairie points 2015-16} \\
	 d4 & {\em Circa 1995 Landcover of the Prairies}\\
	 d5 & {\em Mandatory Food Inspection List}\\
	 d6 & {\em Canadian Food Inspection Agency (CFIA) Fish List}\\
	 d7 & {\em Wholesale trade, sales by trade group}\\
	 d8 & {\em Historical releases of merchandise imports and exports}\\
	 d9 & {\em Immigrant income by period of immigration, Canada}\\
	 d10 & {\em Historical statistics, population and immigrant arrivals}\\
	 \hline
\end{tabular*}
\end{table}
\begin{figure}[h]
\centering
\includegraphics[width=\linewidth]{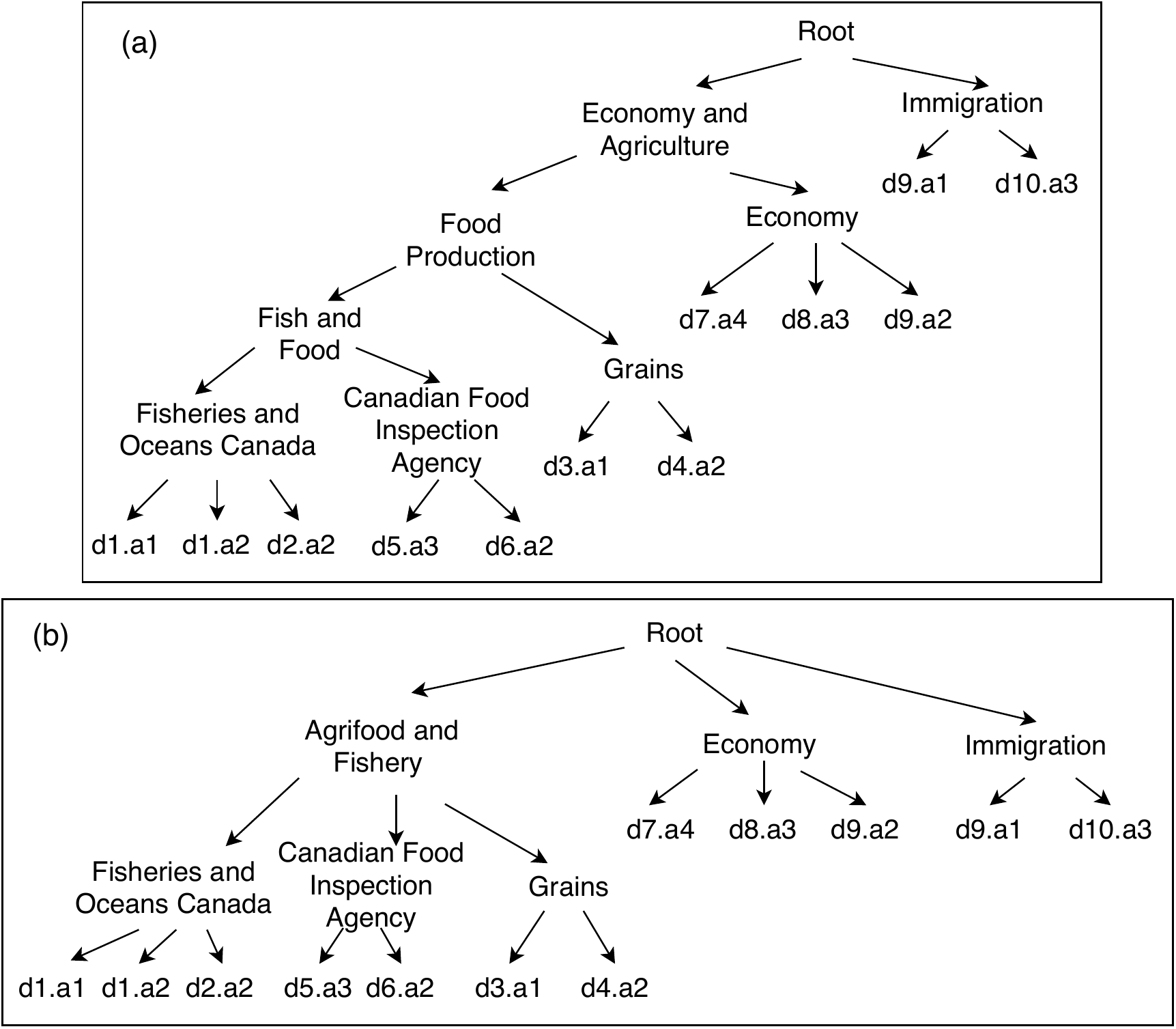}
	\caption{(a) Deep and (b) Effective Organization.}
\label{fig:orgexamples}
\end{figure}
\begin{example}
	\label{ex:runningex}
	Consider the (albeit small) collection of tables 
	from a data lake of open data tables (Table~\ref{tbl:examplerepo}).
A table can be multi-faceted and attributes in a table can be about different topics. 
One way to expose this data to a user is through a flat structure 
of attributes of these tables. 
A user can browse the data (and any associated metadata) and select
data of interest.  If the number of tables and attributes is large, it would be
more efficient to provide an organization over attributes.  
Suppose the tables are organized in the DAG of Figure~\ref{fig:orgexamples}(a).
The label of a non-leaf node in this organization summarizes the content 
of the attributes in the subgraph of the node. 
Suppose a user is interested in the topic of food inspection. 
Using this organization, 
at each step of navigation, they have to choose between only two nodes. 
The first two choices seem clear, {\em Economy and Agriculture} and 
{\em Food Production} seem more relevant to the user than their 
alternatives ({\em Immigration} and {\em Economy}). 
However, having a small branching factor in this organization results 
in nodes that may be misleading. 
For example, it may not be clear if there is any inspection data under the node {\em Fish and Food} or under {\em Grains}. 
This is due to the large heterogeneity of attributes 
(like {\em Oysters} and {\em Grain Elevators}) in the organization below the {\em Fish and Food} node. 
The organization in Figure~\ref{fig:orgexamples}(b) addresses this problem by organizing  
attributes of {\em Grains}, {\em Food Inspection}, and {\em Fisheries} at the same
level.   
Note that this organization has a higher branching factor, but the choices
are more distinct at each node. 
\end{example}

\subsection{Contributions}

We define an {\em organization} as a DAG on nodes consisting of subsets of attributes in a data lake.  
We propose a navigation model on 
an organization which models the user experience during discovery.
We define when an organization is
optimal in that it is
best suited to help a user find any attribute
of interest in a small number of navigation steps without having to consider an excessive number 
of choices or choices that are indistinct.

We make the following contributions.
\squishlist
	\item We propose to model navigation as a Markov model, which computes 
		the probability of discovering a table that is relevant to a topic of interest. 
		We define the {\em data lake organization problem} as the problem of  
		finding an organization that maximizes the expected probability of discovering lake tables, 
		namely the effectiveness of the organization. 		
	\item We propose a local search algorithm that approximates an optimal organization.  
		Particularly, we consider the organization problem as a structure optimization problem 
		in which we explore subsets of the power-set lattice of organizations. 
   \item To reduce the complexity of search, 
		the local search algorithm leverages the lake metadata when it is available. 
		We propose an approximate and efficient way of comparing organizations during search and 
		provide an upper bound for the error of this approximation. 
	\item We propose a metadata enrichment algorithm that effectively bootstraps any existing metadata. 
		Our experiments show that metadata can be transferred across data lakes. 
	\item Our experiments show that organizations are able to optimize navigation based on lake data distribution 
		and outperform a hand-curated taxonomy. 
		We show that the organization constructed by our algorithm outperforms a baseline and an existing linkage graph on a real data lake containing open government data. 
	\item Through a user-study comparing navigation to
                  keyword search, we show that while there is no statistical difference
                in the number of relevant tables found, 
		navigation can help  users find a more
		diverse 
		set of tables than keyword search. 
		In addition, in our study, 
42\% preferred the use of navigation over keyword search (the others preferred search),
suggesting these are complementary and both useful modalities for dataset discovery in data lakes. 
\squishend

\section{Foundations}
\label{sec:model}

We describe a probabilistic model of how users navigate a data lake. 
We envision scenarios in which users
interactively navigate through topics in order to locate relevant tables.
The topics are derived and organized based on 
attributes of tables.  
Our model captures user elements such as the cognitive effort
associated with multiple choices and the increasing complexity of prolonged
navigational steps. 
We then define the data lake organization problem as the 
optimization problem of finding the most effective organization for a data lake.  

\subsection{Organization}
\label{sec:datalakeorg}

Let $\mathcal{T}$ be the set of all tables in a data lake.  
Each table $T\in\mathcal{T}$ consists of a set of attributes, $\mathrm{attr}(T)$. 
Let $\mathcal{A}$ be the set of all attributes,
$\mathcal{A} = \bigcup\{\mathrm{attr}(T):T\in\mathcal{T}\}$ in a data lake.  Each attribute $A$ has
a set of values that we call a domain and denote by $\dom(A)$. 
An organization $\mathcal{O} = (V, E)$ is a DAG. 
Let $\ch(.)$ be the child relation mapping a node to its children, 
and $\parent(.)$ be the parent relation mapping a node to its parents.
A node $s\in V$ is a leaf if $\ch(s)=\emptyset$, otherwise $s$
is an interior node in $\mathcal{O}$.  Every leaf node $s$ of $\mathcal{O}$
corresponds to a distinct attribute $A_s \in \mathcal{A}$.  
Each interior node $s$ corresponds to a set of attributes $D_s\subseteq\mathcal{A}$. 
If $c \in ch(s)$, then $D_c \subseteq D_s$, 
we call this the {\em inclusion} property,
and $D_s = \bigcup_{c \in ch(s) } D_c$. 
We denote the domain of a state $s$ by $\dom(s)$ which is $\dom(A_s)$ when $s$ is a leaf node and 
$\bigcup_{A\in D_s}\dom(A)$ otherwise.

\subsection{Navigation Model}
\label{sec:navigation}

We model a user's experience during discovery using an organization $\mathcal{O}$ as a Markov model 
where states are nodes and transitions are edges.  
We will use the terms state and node interchangably. 
Because of the inclusion property, 
transitions from a state filter out some of the attributes of the state. 
Users select a transition at each step and 
discovery stops once they reach a leaf node. 
To define the effectiveness of an organization, we define a
  user's intent using a query topic $X$ modeled as a set of values. 
Starting at the root node,
a user navigates through sets of attributes (states)
ideally finding attributes of interest. 
\rjmX{In our definition of effectiveness, we assume }
the probability of a user's transition from $s$ to $c\in\ch(s)$
is determined by the similarity between $X$ and the values of the
attributes in $D_c$. 

\begin{example}
	\label{ex:runningex2} 
Returning to Example~\ref{ex:runningex}, to evaluate (and compare)
the effectiveness of the two organizations in
Figure~\ref{fig:orgexamples} for a specific user topic like $X =$ "food
inspection" we can compute the similarity of this topic
to the values in the attributes of the various nodes in each organization.  
\end{example}

Note that when an organization is being used, we do not know
  $X$.  Rather we are assuming that a user performs navigation in a
  way that they choose nodes that are closest to their (unknown to us)
  intended topic query.  The concept of a user's query topic is
  used only to build a good organization.   We 
  define an optimization problem where we build an organization that
  maximizes the expected probability of finding any 
  table in the data lake by finding any of its attributes 
  (assuming a user could potentially {\em have in mind} any attribute
  in the lake).  In other words, the set of query topics we
  optimize for is the set of attributes in the lake. 

\subsection{Transition Model}
\label{sec:transition}

We define the transition probability of $P(\child |s,X,\mathcal{O})$ as the probability that
the user will choose the next state as \child\ if they are at the state $s$.  The
probability should be correlated to the similarity between 
$\dom(\child)$ and $X$. 
Let $\kappa(c, X)$ be a similarity metric between $\dom(\child)$ and $X$.
The transition probability is as follows. 

\begin{equation}
	P(\child |s,X,\mathcal{O}) = \frac
	{e^{\frac{\gamma}{|\ch(s)|} . \kappa(c, X)}}
    {\sum_{t\in ch(s)}e^{\frac{\gamma}{|\ch(s)|} . \kappa(t, X)}}
\label{eq:transprob}
\end{equation}

    The constant $\gamma$ is a hyper parameter of our model.  It must be a
        strictly positive number.
    The term $|\ch(s)|$ is a penalty factor to avoid having nodes with too
        many children.  The impact of the high similarity of a state to
        $X$ diminishes when a state has a large branching factor. 

A discovery sequence is a path, $r=s_1, \ldots, s_k$ 
where $s_i \in ch(s_{i-1})$ for $1 < i \leq k$. 
A state in $\mathcal{O}$ is reached through a discovery sequence.  
The Markov property says that the probability of transitioning 
to a state is only dependent on its parent. 
Thus, the probability of reaching state $s_k$ through a discovery sequence $r=s_1, \ldots, s_k$, 
while searching for $X$ is defined as follows. 
\begin{equation}
	P(s_k|r,X,\mathcal{O}) = \prod_{i=1}^{k}P(s_i|s_{i-1},X,\mathcal{O})
\label{eq:reachprob}
\end{equation}
In this model, a user makes transition choices only based on the current state 
and the similarity of their query topic $X$ to each of the child
states. 
Note that the model naturally penalizes long sequences. 
Since an organization is a DAG, a state can be reached by multiple discovery sequences. 
The probability of reaching a state $s$ in $\mathcal{O}$ while searching for  
$X$ is as follows. 
\begin{equation}
    P(s|X,\mathcal{O}) = \sum_{r\in Paths(s)}P(s|r,X,\mathcal{O})
	\label{eq:pathprob}
\end{equation}
where $Paths(s)$ is the set of all discovery sequences in
$\mathcal{O}$ that reach $s$ from the root. 
Additionally, the probability of reaching a state can be evaluated incrementally.  
\begin{equation}
P(s|X,\mathcal{O}) = \sum_{p \in par(s)} P(s| p, X,\mathcal{O})P(p | X,\mathcal{O})
\label{eq:reachparents}
\end{equation}

\begin{definition} 
	The {\em discovery probability} of an attribute $A$ in organization $\mathcal{O}$ 
	is defined as $P(s|A, \mathcal{O})$, where $s$ is a leaf node. 
    We denote the discovery probability of $A$ as $P(A|\mathcal{O})$.
\label{def:discoveryprob}
\end{definition}

\subsection{Organization Discovery Problem}
\label{sec:orgprob}

A table $\attD$ is discovered in an organization $\mathcal{O}$ by discovering any of its attributes. 

\begin{definition}
\label{dfn:orgEfficiency}
For a single table $\attD$, we define the {\em discovery probability
  of a table} as
\begin{equation}
	P(\attD|\mathcal{O}) = 1-\prod_{\attA\in \attD}(1-P(\attA|\mathcal{O}))
\label{eq:tableprob}
\end{equation}
For a set of tables $\mathcal{T}$ the {\em organization
  effectiveness} is
\begin{equation}
     P(\mathcal{T}|\mathcal{O}) = \frac{1}{|\mathcal{T}|}\sum_{\attD
       \in \mathcal{T}}P(\attD|\mathcal{O})
\label{eq:orgeff}
\end{equation}
\end{definition}

For a data lake, we define the data lake organization problem as
finding an organization that has the highest organization
effectiveness over all tables in the lake.

\begin{definition}
\label{def:orgprob} 
{\em Data Lake Organization Problem.}
	Given a set of tables $\mathcal{T}$ in a data lake, the organization problem is to find 
	an organization $\Hat{\mathcal{O}}$ such that:  
\begin{equation}
	\Hat{\mathcal{O}} = \argmax_{\mathcal{O}} P(\mathcal{T}|\mathcal{O})
\label{eq:orgprob}
\end{equation}
\end{definition}

We remark that in this model we do not assume the availability of any query workloads. 
Since our model uses a standard Markov model, we can apply existing incremental model estimation 
techniques~\cite{DBLP:journals/isci/KhreichGMS12} to maintain and update the transition probabilities 
as behavior logs and workload patterns become available 
through the use of an organization by users. 

\subsection{Multi-dimensional Organizations} 
\label{sec:multidim} 

Given the heterogeneity and massive size of data lakes, it may
  be advantagous to perform an initial grouping or clustering of
  tables and then build an organization on each cluster. 
\begin{example}
\label{ex:multiD}
Consider again the tables from Example~\ref{ex:runningex} which come
from real government open data.  To build an organization over 
a large open data lake, we can first cluster tables (e.g., using ACSDb which
clusters tables based on their
schemas~\cite{DBLP:journals/pvldb/CafarellaHWWZ08} or using a
	clustering over the attribute values or other features of a table)~\cite{DBLP:journals/pvldb/CafarellaHK09}. 
	This would define
(possibly overlapping) groups of tables, perhaps a group on
Immigration data and another on Environmental data over which we may
be able to build more effective organizations. 
\end{example}
If we have a grouping of  $\mathcal{A}$ into $k$ (possibly overlapping)
  groups or dimensions, $G_1, \ldots, G_k$, then we can discover an
  organization over each and use them collectively for navigation. 
We define a $k$-dimensional organization $\mathcal{M}$ for a data lake $\mathcal{T}$ as 
a set of organizations $\{\mathcal{O}_1, \ldots, \mathcal{O}_k\}$, 
such that attributes of each table in $\mathcal{T}$ 
is organized in at least one organization and 
$\mathcal{O}_i$ is the most effective organization for $G_i$.  
We define the probability of discovering table $\attD$ in $\mathcal{M}$, 
as the probability of discovering $\attD$ in any of dimensions of $\mathcal{M}$. 
\begin{equation}
	P(\attD|\mathcal{M}) = 1-\prod_{\mathcal{O}_i\in \mathcal{M}}(1-P(\attD|\mathcal{O}_i))
\label{eq:multidim-prob}
\end{equation}

\section{Constructing Organizations}
\label{sec:algorithm}

Given the abstract model of the previous section, we now present
  a specific instantiation of the model that is suited for real data
  lakes.  
  We justify our design
choices using lessons learned from search-based table discovery.  We
then consider the metadata that is often available in data lakes,
specifically table-level tags, and explain how this metadata (if
available) can be exploited for
navigation.  Finally, we present a local search algorithm for building an
approximate solution for the data lake organization problem.

\subsection{Attribute and State Representation}
\label{sec:stateattrep}

We have chosen to construct organizations over the text attributes of data lakes. 
This is based on the observation that although a small percentage of attributes in a lake are text attributes 
(26\% for a Socrata open data lake that we use in our experiments), 
the majority of tables (92\%) have at least one text attribute.  
We have found that similarity between numerical attributes
  (measured by set overlap or Jaccard) can be very misleading as
  attributes that are semantically unrelated can be very similar and
  semantically related attributes can be very dissimilar.  
  Hence, to use numerical attributes one would first need to
  understand their semantics.  Work is emerging on how to do
  this~\cite{DBLP:conf/cikm/IbrahimRW16,DBLP:conf/kdd/HulsebosHBZSKDH19,DBLP:conf/icde/IbrahimRWZ19}.

Since an organization is used for the exploration of heterogeneous lakes,
we are interested in the semantic similarity of values.
To capture semantics, a domain can be canonicalized by a set of
relevant class labels  from a knowledge base~\cite{Limaye:2010:ASW,venetis2011recovering,DBLP:conf/kdd/HulsebosHBZSKDH19}. 
We have found that open knowledge bases have relatively low coverage on data lakes~\cite{NargesianZPM18}.
Hence, we follow an approach which has proven to be successful in
search-based table discovery which is to represent attributes by their collective word embedding 
vectors~\cite{NargesianZPM18}.

Each data value $v$ is represented by an 
embedding vector $\vec{v}$ such that values that are more likely to share the
same context have embedding 
vectors that are close in embedding space according to an Euclidean or
angular distance measure~\cite{Mikolov:2013}. 
Attribute $A$ can be represented by a {\em topic vector}, $\mu_A$, 
which is the sample mean of the population $\{\vec{v} | v\in\dom(A)\}$~\cite{NargesianZPM18}.  
In organization construction, we also represent $X$ by $\mu_X$. 
\begin{definition}
\label{def:state}
	State $s$ is represented with a {\em topic vector}, $\mu_s$, 
which is the sample mean of the population $\{\vec{v} | v\in\dom(s)\}$. 
\end{definition}

If a sufficient number of values have word embedding representatives,  
the topic vectors of attributes are good indicators of attributes' topics. 
We use the word embeddings of fastText database~\cite{joulin:2016}, 
which on average covers 70\% of the values 
in the text attributes in the
datasets used in our experiments.  
In organization construction, to 
evaluate the transition probability of navigating to state $\child$ 
we choose $\kappa(\child, X)$ to be the Cosine similarity between $\mu_{\child}$ and $\mu_X$. 
Since a parent subsumes the attributes in its children, 
the Cosine similarity satisfies a monotonicity property of 
$\kappa(X, \child) > \kappa(X, s)$, where $\child \in ch(s)$. 
However, the monotonicity
property does not necessarily hold for the transition probabilities.  This is
because $P(\child | s, X, O)$ is normalized with all children of parent $s$. 

\subsection{State Space Construction}
\label{sec:datalakes}

\paragraph{Metadata in Data Lakes}
Tables in lakes are sometimes accompanied by metadata 
hand-curated by the publishers and creators of data.  
In enterprise lakes, metadata may come from design documentation from
which tags or topics can be extracted.  For
open data, the standard APIs used to publish data include tags or
keywords~\cite{DBLP:journals/debu/MillerNZCPA18}.  
In mass-collaboration datasets, like web tables, contextual
 information can be used to extract metadata~\cite{hassanzadeh2015understanding}.

\paragraph{State Construction}
If metadata is available, it can be distilled into tags 
  (e.g., keywords, concepts, or entities). 
We associate these table-level tags with every attribute in the table. 
In an organization, the leaves still have a single attribute,
  but the immediate parent of a leaf is a state containing all
  attributes with a given (single) tag.  We call these parents {\em
    tag states}. 
Building organizations on tags reduces the number of possible states and the
size of the organization while still having meaningful nodes 
  that represent a set of attributes with a common tag. 
\begin{example}
\label{ex:tags}
In the organization of Figure~\ref{fig:orgexamples} (b) we
have actually used real tags from open data to label the
internal nodes (with the exception of {\em Agrifood and Fishery} which
is our space saving representation of a node representing the three real
tags:  {\em Canadian Food Inspection Agency}, {\em Fisheries and
	Oceans Canada} and {\em Grains}).  Instead of building an
organization over the twelve attributes, we can build a smaller
organization over the five tags:   {\em Canadian Food Inspection
  Agency}, {\em Fisheries and   Oceans Canada}, {\em Grains}, {\em
  Economy}, and {\em Immigration}. 
\end{example}
\begin{definition}
\label{def:tagState}
	Let $\data(t)$ be the relation mapping a tag $t$ in a lake metadata 
	to the set of its associated attributes. 
	Suppose $M_s$ is the set of tags in a tag state $s$.   
	Thus, the set of attributes in $s$ is $D_s = \bigcup_{t\in M_s}\data(t)$.  
	A tag state $s$ is represented with a {\em topic vector}, $\mu_s$,  
which is the sample mean of the population $\{\vec{v} | v\in\dom(A), A\in D_s\}$. 
\end{definition}

\paragraph{Flat Organization: A Baseline} 
In the organization, we consider the leaves to be
states with one attribute.
However, now the parent of a leaf node is associated to only one tag. 
This means that the last two levels of a hierarchy are fixed 
and an organization is constructed over states with a single tag. 
If instead of discovery, we place a single root node over such 
states, we get a {\em flat organization} that we can use as a
baseline.  This is a reasonable baseline as it is conceptually, the
navigation structure supported by many open data APIs that permit
programmatic access to  open data resources.  These APIs permit
retrieval of tables by tag.  

\label{sec:metadata}
\paragraph{Enriching Metadata}
Building organizations on tags reduces organization discovery cost. 
However, the tags coming from the metadata may be
incomplete (some attributes may have no tags).
Moreover, the schema and vocabulary of metadata across data
originating from different sources may be
inconsistent which can lead to disconnected organizations. 
We propose to transfer tags across data lakes 
such that data lakes with no (or little) metadata are augmented with
the tags from other data lakes.  
To achieve this,   
we build binary classifiers, one per tag, which predict the association of attributes to 
the corresponding tags. 
The classifiers are trained on the topic vector, $\mu_A$, of attributes.   
Attributes that are associated to a tag are the positive training samples for the tag's classifier 
and the remaining attributes are the negative samples. 
This results in imbalanced training data for large lakes.  
To overcome this problem, we only use a subset of negative samples for training. 
We consider the ratio of positive and negative samples as 
a hyperparameter.

\subsection{Local Search Algorithm}

\newcommand{\algName}{\mbox{\tt Organize}}

\begin{algorithm}[t]
	\caption{Algorithm \algName}
    	\label{alg:organize}
    	\begin{algorithmic}[1]
	\Procedure{\textsc{\algName}}{$\mathcal{T}$}
		\State $mod\_ops = [\textsc{Delete\_Parent}, \textsc{Add\_Parent}]$
		\State $\mathcal{O} \gets \textsc{Init\_Org}(\mathcal{T})$, $p \gets \textsc{eval}(\mathcal{O})$\label{line:init}
		\While{$\neg termination\_cond$}\label{line:terminate}
		\State $state \gets \textsc{state\_to\_modify}(\mathcal{O})$\label{line:stateSelect}
		\State $\mathcal{O}^\prime \gets \textsc{choose\_apply\_op}(\mathcal{O}, state, mod\_ops)$\label{line:opSelect}
		\If{$\textsc{accept}(\textsc{eval}(\mathcal{O}^\prime), p)$}\label{line:accept}
		\State $\mathcal{O} \gets \mathcal{O}^\prime, p \gets p^\prime$
		\EndIf
	\EndWhile
    	\State \textbf{return} $\mathcal{O}$
    	\EndProcedure
    	\end{algorithmic}
\end{algorithm}

Our local seach algorithm, \algName, is outlined in 
Algorithm~\ref{alg:organize}. 
It begins with an initial organization (Line~\ref{line:init}).  
At each step, the algorithm proposes a {\em modification} to the current organization $\mathcal{O}$ 
which leads to a new organization $\mathcal{O}^\prime$ 
(Line~\ref{line:opSelect}). 
If the new organization is closer to a solution for the {\em
    Data Lake Organization Problem} (Definition~\ref{def:orgprob}), it
  is   accepted as the new organization, otherwise it is accepted
  (Line~\ref{line:accept}) with a
  propability that is a ratio of the effectiveness, namely~\cite{FriedmanK03}: 
\begin{equation}
min[1, \frac{P(\mathcal{T}|\mathcal{O}^\prime)}{P(\mathcal{T}|\mathcal{O})}]
\end{equation}
The algorithm terminates (Line~\ref{line:terminate}) once the effectiveness of an
organization reaches a plateau.
In our experiments, we terminate when the expected probability has not improved significantly
for the last 50 iterations. 

We heuristically try to maximize the effectiveness of an organization 
by making its states highly reachable. 
We use Equation~\ref{eq:reachparents} to evaluate the probability of reaching a 
state when searching for attribute $\attA$ and  
define the overall {\em reachability probability} of a state as follows. 
\begin{equation}
P(s|\mathcal{O}) = \frac{1}{|\attA\in\mathcal{T}|} \sum_{\attA\in \mathcal{T}}P(s|\attA,\mathcal{O})
\end{equation}
Starting from an initial organization, 
the search algorithm performs downward traversals from the root and 
proposes a modification on the organization for states in each level of
the organization ordered 
from lowest reachability probability to highest 
(Line~\ref{line:stateSelect} and~\ref{line:opSelect}). 
A state is in level $l$ 
if the length of the shortest discovery paths 
from root to the state is $l$. 
We restrict our choices of a new organization 
at each search step to those created by two operations. 

\begin{figure*}[t]
\centering
	\subfloat[\label{fig:add-parent}]{\includegraphics[width=0.25\linewidth]{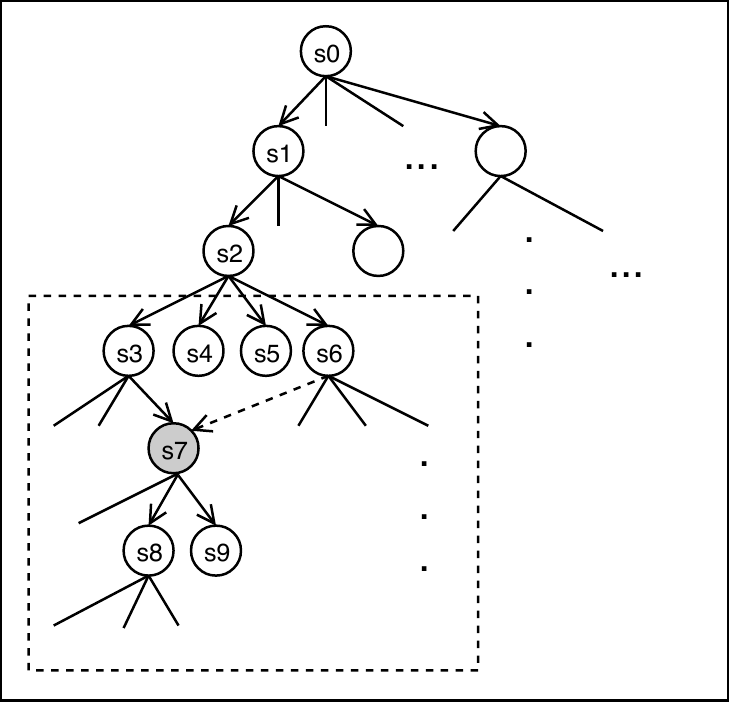}}
\hspace{0.02cm}
	\centering
	\subfloat[\label{fig:reduce-height-before}]{\includegraphics[width=0.25\linewidth]{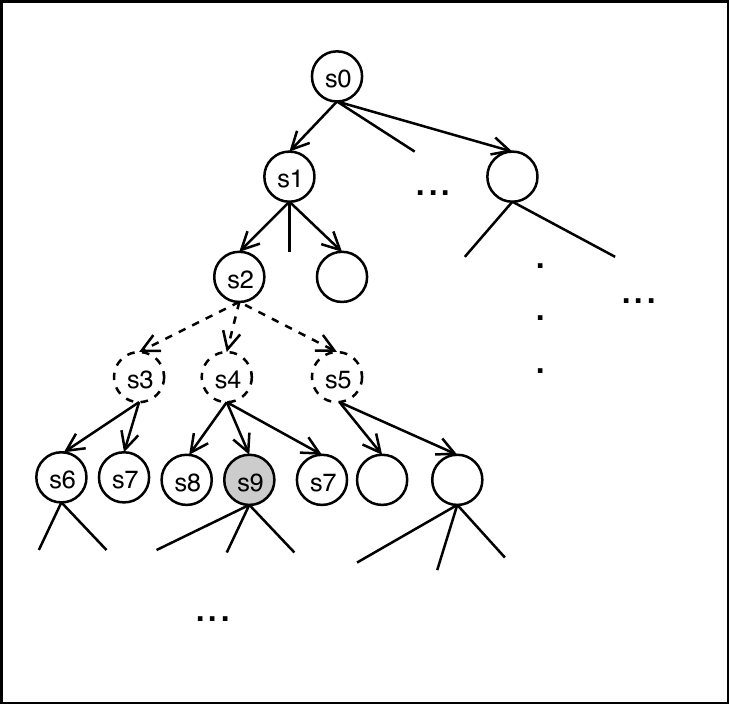}}
	\hspace{0.02cm}
	\centering
	\subfloat[\label{fig:reduce-height-after}]{\includegraphics[width=0.25\linewidth]{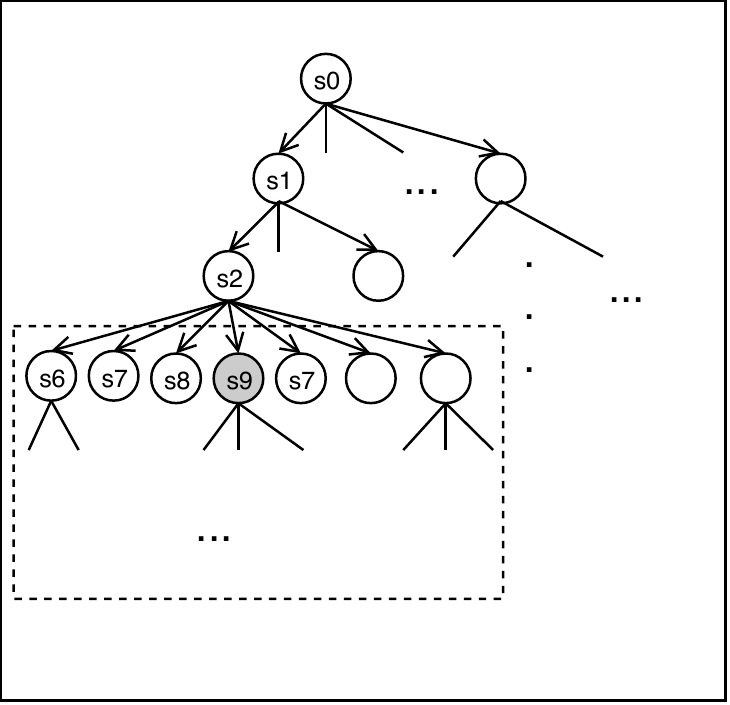}}
\caption{
	(a) Before and After Applying \textsc{Add\_Parent} Operation, (b) Before and (c) After Applying \textsc{Delete\_Parent} Operation.}
\label{fig:fix-ops}
\end{figure*}

\paragraph*{Operation I: Adding Parent} 
Given a state $s$ with low reachability probability, one reason for this
  may be that it is one child amongst many of its current parent, or that
  it is indistinguishable from a sibling.  We can remedy either of
  these by adding a new parent for $s$. 
Suppose that the search algorithm chooses to modify the organization with respect to state $s$. 
Recall that Equation~\ref{eq:reachparents} indicates that the probability of reaching a state 
increases as it is connected to more parent states. 
Suppose $s$ is at level $l$ of organization $\mathcal{O}$. 
The algorithm finds the state, called $n$, at level $l-1$ of $\mathcal{O}$ such that it is not a parent of $s$ and 
has the highest reachability probability among the states at $l-1$. 
To satisfy the inclusion property, we update node $n$ and its ancestors 
to contain the attributes in $s$, $D_s$. 
To avoid generating cycles in the organization, 
the algorithm makes sure that none of the children of $s$ is chosen as a new parent. 
State $n$ is added to the organization as a new parent of $s$. 
Figure~\ref{fig:add-parent} shows an example of {\tt ADD\_PARENT} operation. 
{\tt ADD\_PARENT} potentially increases
the reachability probability of a state by adding more discovery paths  
ending at that state, at the cost of increasing the branching factor.  

\begin{example}
For example, the attribute {\tt d6.a2} from table 
{\em Canadian Food Inspection Agency (CFIA) Fish List} 
in the organization of  Figure~\ref{fig:orgexamples}(b) is reachable 
from the node with the tag {\em Canadian Food Inspection Agency}. 
This attribute is also related to the node with the tag {\em Fisheries and Oceans Canada} 
and can be discovered from that node. 
The algorithm can decide to add an edge from 
{\em Fisheries and Oceans Canada} to {\tt d6.a2} and make this state its second parent. 
\end{example}
\paragraph*{Operation II: Deleting Parent} 
\rjmX{Another reason a state can have low reachability is}
  that its parent has low reachability probability and we should perhaps
  remove a parent. 
Reducing the length of paths from the root to state $s$ is a second way to boost the 
reachability probability of $s$. 
The operation eliminates the least reachable parent of $s$, called $r$. 
To reduce the height of $\mathcal{O}$, the operation eliminates all siblings of $r$ 
except the ones with one tag. 
Then, it  
connects the children of each eliminated state to its parents. 
Figures~\ref{fig:reduce-height-before} and~\ref{fig:reduce-height-after} show an example of applying 
this operation on an organization. 
This makes the length of paths to $s$ smaller which boosts the reachability probability of $s$. 
However, replacing the children of a state by all its grandchildren 
increases the branching factor of the state, thus, 
decreasing the transition probabilities from that state. 

\begin{example} 
For example, the algorithm can decide to eliminate the state {\em Economy and Agriculture}  
in the organization of Figure~\ref{fig:orgexamples}(a) and 
connect its children directly to the root. 
\end{example}

In Algorithm~\ref{alg:organize}, {\tt STATE\_TO\_MODIFY} (Line~\ref{line:stateSelect}) orders
  states by level starting at level 1 and within a level by
  reachability (lowest to highest) and returns the next state each
  time it is called.  {\tt CHOOSE\_APPLY\_OP} picks the operator that 
  when applied 
creates the most effective organization $\mathcal{O}^\prime$. 
  Everytime a new organization is chosen, 
  {\tt STATE\_TO\_MODIFY} may need to reorder states
  appropriately.  

\paragraph*{Initial organization} (Line~\ref{line:init}) 
\label{sec:initorg}
The initial organization may be any organization that satisfies the
inclusion property of attributes of states. 
For example, the initial organization can be the DAG defined based on a hierarchical clustering of the 
tags of a data lake. 
 
\section{Scaling Organization Search}
\label{sec:scale}

The search algorithm makes local changes to an existing organization
by applying operations. 
An operation is successful if it increases 
the effectiveness of an organization.  
The evaluation of the effectiveness 
(Equation~\ref{eq:orgprob})
involves computing the discovery probability for all attributes  
which requires evaluating the probability of reaching the states along the paths to an attribute 
(Equation~\ref{eq:reachparents}). 
The organization graph can have a large number of states, especially at the initialization phase.
To improve the search efficiency, we first identify the subset
  of states and attributes whose  discovery probabilities may be changed by an
  operation and second we approximate the new discovery probabilities
  using a set of attribute representatives. 

\subsection{States Affected by an Operation}
\label{sec:lazyeval}

At each search iteration we only re-evaluate the discovery probability
of the states  and attributes 
which are affected by the local change. 
Upon applying \textsc{Delete\_Parent} on a state, the transition probabilities  
from its grandparent to its grandchildren are changed and consequently all states reachable from the grandparent. 
However, the discovery probability of states
that are not reachable from the grandparent 
remain intact. 
Therefore, for \textsc{Delete\_Parent}, we only re-evaluate 
the discovery probability of the states in the sub-graph rooted by the grandparent and 
only for attributes associated to the leaves of the sub-graph. 

The \textsc{Add\_Parent} operation impacts the organization more broadly. 
Adding a new parent to a state $s$ changes the discovery probability of $s$ and all states that are reachable from $s$. 
Furthermore, the parent state and consequently its ancestors are updated 
to satisfy the inclusion property of states. 
Suppose the parent itself has only one parent. 
The change of states propagates to all states up to the lowest common
ancestor (LCA) of $s$  
and its parent-to-be before adding the transition to the organization. 
If the parent-to-be has multiple parents the change
needs to be
propagated to other subgraphs. 
To identify the part of the organization that requires re-evaluation, 
we iteratively compute the LCA of $s$  
and each of the parents of its parent-to-be. 
All states in the sub-graph of the LCA require re-evaluation. 

\subsection{Approximating Discovery Probability}  
\label{sec:approx}

Focusing on states
that are affected by operations reduces 
the complexity of the exact evaluation of an organization. 
To further speed up search, 
we evaluate an organization on a small number of attribute representatives. 
that each summarizes a set of attributes. 
The discovery probability of each representative approximates the
discovery probability of 
its corresponding attributes. 
We assume a one-to-one mapping between representatives and a partitioning of attributes. 
Suppose $\rho$ is a representative for a set of attributes $D_{\rho} = \{\attA_1, \ldots, \attA_m\}$. 
We approximate $P(\attA_i|\mathcal{O})$, $\attA_i\in D_{\rho}$  
with $P(\rho|\mathcal{O})$. 
The choice and the number of representatives impact the error of this approximation. 

Now, we describe an upper bound for the error of the approximation. 
Recall that the discovery probability 
of a leaf state is the product of transition probabilities   
along the path from root to the state.  
To determine the error that a representative introduces to the discovery probability of an attribute,  
we first define an upper bound on the error incurred by 
using representatives in transition probabilities. 
We show that the error of transition probability from $m$ to $s$ is bounded by a fraction of the actual  
transition probability which is correlated with the similarity of the representative to the attribute. 
Recall the transition probability from Equation~\ref{eq:transprob}. 
For brevity, 
we assume $\gamma^\prime=\frac{\gamma}{|ch(m)|}$. 
\begin{equation}
P(s_i|m, \attA,\mathcal{O}) = \frac{e^{\gamma^\prime\kappa(\attA, s_i)}}{\sum_{s_j\in ch(m)}e^{\gamma^\prime \kappa(\attA, s_j)}}
\label{eq:brevtransprob}
\end{equation}
Suppose $\delta$ is the distance metric of the metric $\kappa$, 
which is $\delta(a,b)=1-\kappa(a,b)$. 
From the triangle property, it follows that:  
\begin{equation}
	\delta(s_i, \rho_k) \leq \delta(s_i, \attA) + \delta(\attA, \rho)
\end{equation}
Evaluating $P(s_i|m, \attA,\mathcal{O})$ and $P(s_i|m, \rho,\mathcal{O})$ requires 
computing $\kappa(\attA, s_j)$ and $\kappa(\rho, s_j)$ on $\ch(m)$.  
We rewrite the triangle property as follows:   
\begin{equation}
	1 - \kappa(s_i, \rho) \leq 1 - \kappa(s_i, \attA) + 1 - \kappa(\attA, \rho)
\end{equation}
Therefore, the upper bound of $\kappa(s_i, \attA)$ is defined as follows. 
\begin{equation}
	\kappa(s_i, \attA) \leq \kappa(s_i, \rho) - \kappa(\attA, \rho) + 1
\end{equation}
We also have the following. 
\begin{equation}
	0 \leq \kappa(s_i, \attA) - \kappa(s_i, \rho) \leq 1 - \kappa(\attA, \rho) 
\label{eq:simbounds}
\end{equation}

Let $\Delta_i=\kappa(s_i, \attA)-\kappa(s_i, \rho)$. 
Without loss of generality, we assume $\kappa(s_i, \attA) > \kappa(s_i, \rho)$, 
thus $\Delta_i$ is a positive number.  
Now, we can rewrite $\kappa(s_i, \rho)=\kappa(s_i, \attA)-\Delta_i$. 
From Equation~\ref{eq:transprob}, 
we know that $P(s_i|m, \attA,\mathcal{O})$ and $P(s_i|m, \rho,\mathcal{O})$ are monotonically increasing with 
$\kappa(s, \attA)$ and $\kappa(s_i, \rho)$, respectively. 
Therefore, the error of using the representative $\rho$ in approximating 
the probability of transition from $m$ to $s_i$ when looking for $\attA$ 
is as follows: 
\begin{equation}
\epsilon = P(s_i|m, \attA,\mathcal{O}) - P(s_i|m, \rho,\mathcal{O})  
\end{equation}
It follows from the monotonicity property that $\epsilon\geq0$. 
By applying Equation~\ref{eq:brevtransprob} to $\epsilon$, we have the following: 
\begin{equation}
\epsilon = \frac{e^{\gamma^\prime\kappa(\attA, s_i)}}{\sum_{s_j\in ch(m)}e^{\gamma^\prime\kappa(\attA, s_j)}}
-
\frac{e^{\gamma^\prime\kappa(\rho, s_i)}}{\sum_{s_j\in ch(m)}e^{\gamma^\prime\kappa(\rho, s_j)}}
\label{eq:transapprox}
\end{equation}
We rewrite the error by replacing $\kappa(s_i, \rho)$ with $\kappa(s_i, \attA)-\Delta_i$. 
\begin{equation}
\epsilon = \frac{e^{\gamma^\prime\kappa(\attA, s_i)}}{\sum_{s_j\in ch(m)}e^{\gamma^\prime\kappa(\attA, s_j)}}
-
\frac{e^{\gamma^\prime\kappa(s_i, \attA)}.e^{-\gamma^\prime\Delta_i}}{\sum_{s_j\in ch(m)}e^{\gamma^\prime\kappa(s_j, \rho)}.e^{-\gamma^\prime\Delta_j}}
\label{eq:transerror}
\end{equation}
Following from Equation~\ref{eq:simbounds}, we have: 
\begin{equation}
\frac{1}{e^{\gamma^\prime(1 - \kappa(\attA, \rho))}} \leq e^{-\gamma^\prime\Delta_i} \leq 1
\end{equation}
The upper bound of the approximation error of the transition probability is: 
\begin{multline}
\epsilon \leq \frac{e^{\gamma^\prime\kappa(\attA, s_i)}}{\sum_{s_j\in ch(m)}e^{\gamma^\prime\kappa(\attA, s_j)}}
- 
\frac{e^{\gamma^\prime\kappa(\attA, s_i)}}{\sum_{s_j\in ch(m)}e^{\gamma^\prime\kappa(\attA, s_j)}}.\\ 
\frac{1}{e^{\gamma^\prime(1-\kappa(\rho,\attA))}}
\label{eq:transerror}
\end{multline}
The error can be written in terms of the transition probability to a state given an attribute: 
\begin{equation}
\epsilon \leq P(s_i|m, \attA,\mathcal{O}).(1-\frac{1}{e^{\gamma^\prime(1-\kappa(\rho,\attA))}})
\label{eq:transerror}
\end{equation}
Since $e^{\gamma^\prime(1-\kappa(\rho,\attA))} \geq 1$, the error is bounded. 

Assuming the discovery path $r=s_1, \ldots, s_k$, 
the bound of the error of approximating $P(\attA_i|\mathcal{O})$ using $\rho$ is as follows:
\begin{equation}
\epsilon_{r} \leq (\prod_{i=1}^{k}P(s_i|s_{i-1}, \attA,\mathcal{O})).(1-\frac{1}{e^{\gamma^\prime(1-\kappa(\rho,\attA))}})^k
\end{equation}
To minimize the error of approximating $P(s|m, \attA,\mathcal{O})$ considering $\rho$ instead of $\attA$,  
we want to choose $\rho$'s that have high similarity to the attributes they represent, 
while keeping the number of $\rho$'s relatively small. 

\subsection{Dynamic Data Lakes}

Updating data lakes changes the states of its organization and as a result 
the transition probabilities would change. 
Now, the original organization that was optimized might not 
be optimal anymore because of the changes in transition probabilities. 
Suppose upon an update to a data lake, 
a state $s_i$ is updated to $s^\prime_i$. 
If $s_i$ is a state in the optimal organization, 
the error upper bound provided for representatives 
provides a guide for how 
$s^\prime_i$ has diverged from  $s_i$. 
The upper bound of the change 
in transition probabilities when $s_i$ is updated to $s^\prime_i$ is as follows. 
\begin{equation}
\label{eq:orgupdate}
\alpha \leq P(s_i|m,\attA,\mathcal{O}).(1-\frac{1}{e^{\gamma^\prime(1-\kappa(s_i, s^\prime_i)}})
\end{equation}
This error bound can be used to determine when data changes are significant enough to require rebuilding an organization. 

\section{Evaluation}
\label{sec:evaluation}

We first seek to quantify and understand some of the design
  decisions, the efficiency, and the influence of using approximation
  for our approach.  We do this using a small
  synthesized benchmark called {\tt TagCloud} that is designed so that we
  know precisely the best tag per attribute.  Next, using real open
  data, in Section~\ref{sec:comparisons}, we quantify the benefits of
  our approach over 1) an existing ontology (Yago~\cite{Yago})
  for navigation; 2) a flat baseline (where we ask what value is
  provided by the hierarchy we create over just grouping tables by
  tags); and 3) a table linkage graph, called an Enterprise
  Knowledge Graph~\cite{fernandez2018aurum}, to navigate.  Our
  approach uses table-level tags, so we also illustrate that tags from
  a real open data lake can be easily transferred to a different data
  lake with no metadata using a simple classifier
  (Section~\ref{sec:metadataenrich}).  Finally, we present a user
  study in Section~\ref{sec:userstudy}.

\subsection{Datasets} 
\label{sec:datasets}

We begin by describing our real and synthetic datasets which are
summarized in Table~\ref{tbl:datasets}. 

\begin{table}[h]
\small
\centering
	\caption{Experimental Datasets}
	\label{tbl:datasets}
\begin{tabular}{@{} l c c c @{} }
\toprule 
	\textbf{Name} & \textbf{\#Tables} & \textbf{\#Attr} &
                                                            \textbf{\#Tags} \\
\midrule  
Socrata &   7,553 &		50,879 	&	11,083\\
Socrata-1 & 1,000	& 5,511&			 3,002\\
Socrata-2 & 2,175	& 13,861 &	345	\\
Socrata-3 & 2,061	& 16,075 &	346	\\
CKAN	 & 1,000	&	 7,327	&	 0\\
TagCloud & 369 &		 2,651 &		 365\\
YagoCloud & 370 & 		 2,364	&	 500\\
\bottomrule
\end{tabular}
\end{table}

\begin{figure}[t]
    \centering
	\subfloat[\label{fig:tablegroups-dist}]{\includegraphics[width=0.49\columnwidth]{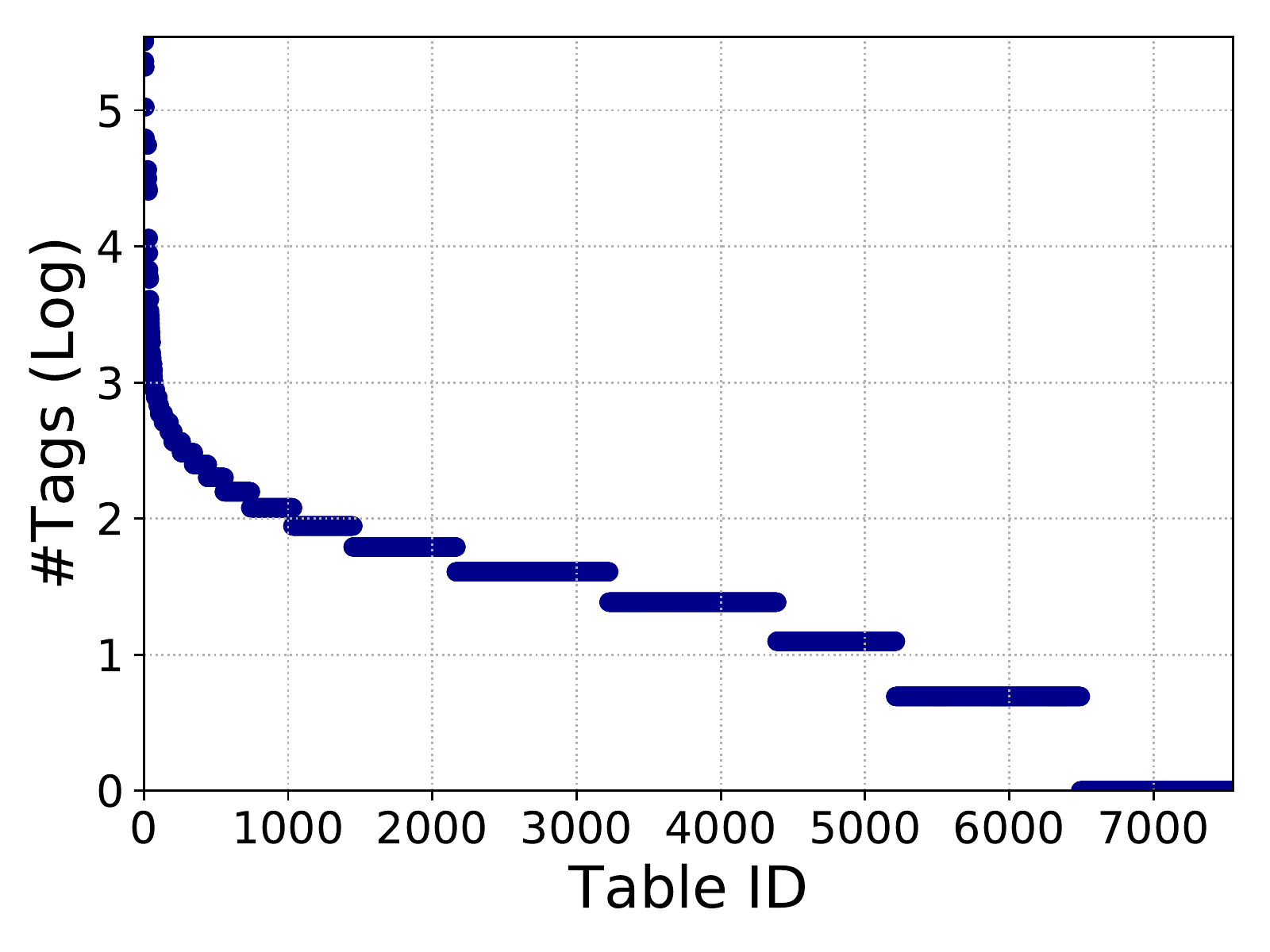}}
\centering
	\subfloat[\label{fig:tableatts-dist}]{\includegraphics[width=0.49\columnwidth]{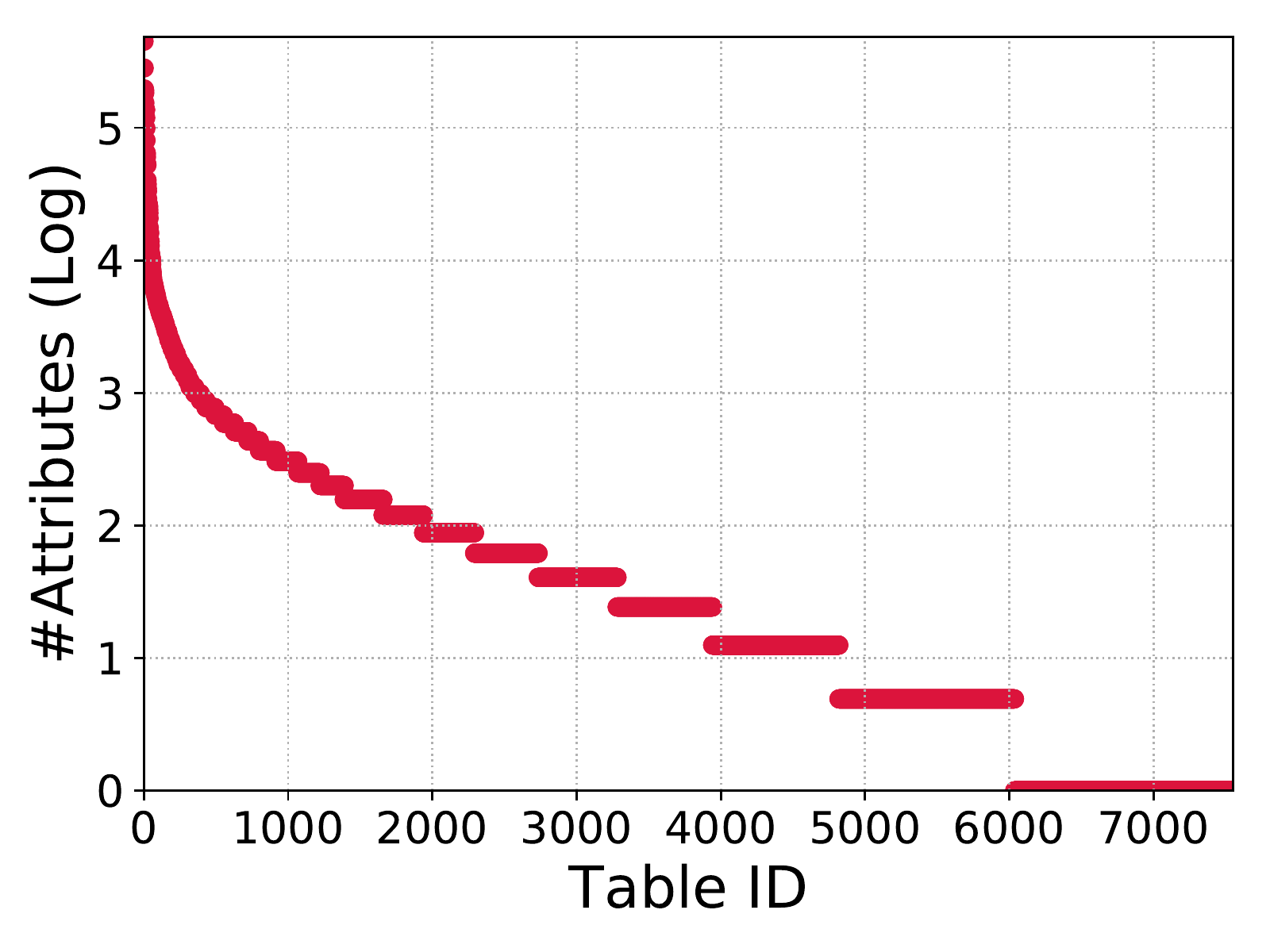}}
\caption{
	Distribution of (a) Tags and (b) Attributes per Table in {\tt Socrata} Data Lake.}
\label{fig:od-stats}
\end{figure}

{\bf Socrata and CKAN Data Lakes - }
For our comparison studies, we used real open data. 
We crawled 7,553 tables with 11,083 tags
from the Socrata API. 
We call this lake {\tt Socrata}. 
It contains 50,879 attributes containing words that have a word embedding. 
In this dataset, a table may
be associated with many tags and attributes inherit the tags of their table. 
We have 264,199 attribute-tag associations. 
The distribution of tags per table and 
attributes per table of {\tt Socrata} lake is plotted in 
Figure~\ref{fig:od-stats}.  
The distribution is skewed with two tables having over 100K tags and
  the majority of the tables having 25 or fewer. 
{\tt Socrata-1} is a random collection of 1,000 tables and 3,002 tags 
from {\tt Socrata} lake that we use in our comparison with the 
Enterprise Knowledge graph.
{\tt Socrata-2} is a collection of 
2,061 tables and 345 tags from {\tt Socrata} and 
{\tt Socrata-3} is a 
collection of 2,175 tables and 346 tags from {\tt Socrata}.  
Note that {\tt Socrata-2} and {\tt Socrata-3} do not share any tags and are used in our user study.    
The {\tt CKAN} data lake is a separate collection of 1,000 tables and 7,327
attributes from the CKAN API. 
For the metadata enrichment experiments, we removed all tags from the
{\tt CKAN} data lake and study how metadata from {\tt Socrata} can be
  transferred to {\tt CKAN} (Section~\ref{sec:metadataenrich}).

{\bf TagCloud benchmark - }
To study the impact of the density of metadata and the
  usefulness of multi-dimensional organizations, we synthesized a dataset where we know exactly the most relevant tag for an attribute.
  Note that in the real open data, tags may be incomplete or
  inconsistent (data can be mislabeled).  We create only a single tag
  per attribute which is actually a disadvantage to our approach that
benefits from more metadata.  The benchmark is small so we can report
both accuracy and speed for the non-approximate version of our
algorithm in comparison to the approximate version that computes
discovery probabilities using attribute representatives. 
We synthesized a collection of 369 tables with 2,651 attributes.
First, we generate tags by choosing a sample of 365
words from the fastText database that 
are not very close according to Cosine similarity. 
The word embeddings of these words are then used to generate attributes 
associated to tags. 
Each attribute in the benchmark is associated to exactly one tag. 
The values of an attribute are samples of a normal distribution 
centered around the word embedding of a tag. 
To sample from the distribution of a tag, we selected the $k$ most similar
words, based on Cosine similarity, 
to the tag, where $k$ is 
the number of values in the attribute 
(a random number between 10 and 1000). 
This guarantees that the distribution of the word embedding of 
attribute values has small variance and 
the topic vector of attributes are close to their tags. 
This artificially guarantees that the states that contain the tag of an attribute 
are similar to the attribute 
and likely have high transition probabilities. 

To emulate the metadata distribution of real data lakes (where
  the number of tags per table and number of attributes per table
  follow Zipfian distributions (Figure~\ref{fig:od-stats})), we generated tables so that
  the number of tags (and therefore attributes) also follows a Zipfian
distribution. 
In the benchmark the number of attributes per table is sampled from [1, 50] 
following a Zipfian distribution.

{\bf YagoCloud benchmark - }
To have a fair comparison with YAGO, we cannot use our real open
  data lakes because YAGO has low coverage over this data. 
Hence, we synthesis a data lake where YAGO has full coverage and
  therefore the class hierarchy of YAGO would be a reasonable alternative for navigation. 
{\tt YagoCloud} is a collection of 370 tables with 2,364 attributes 
that can be organized using the taxonomy of YAGO. 
Like {\tt TagCloud} benchmark, 
the distribution of attributes in tables and 
tags per table in this benchmark  
are created to 
emulate the observed characteristics of our crawl of open data portals (Figure~\ref{fig:od-stats}). 
In YAGO, each entity is associated with one or more {\em types}.  
We consider 500 random leaf types that 
contain the words ``agri'', ``food'', or ``farm'' in their labels.  
These types are equivalent to tags. 
Each attribute in the benchmark is associated to exactly one tag. 
For each attribute of a table, 
we randomly sample its values from the set of entities associated to its tag. 
This guarantees that the attribute is about the type (tag).  
We tokenize the sampled entities into words and 
the word embeddings of these words are then used to generate 
the topic vectors of attributes. 
The number of values in an attribute is a random number between 10 and 1000. 
and the number of attributes per table is sampled from [1, 50] 
following a Zipfian distribution.  

The sub-graph of the YAGO taxonomy that covers all ancestor classes 
of the benchmark types 
is considered as the taxonomy defined on benchmark attributes. 
This taxonomy is a connected and acyclic graph. 
Each class in the taxonomy is equivalent to a non-leaf state of navigation. 
To guarantee the inclusion property in the taxonomy, 
each non-leaf state consists of the tags corresponding to its descendant types. 
The topic vector of an interior state is generated by aggregating the topic vectors of 
its descendant attributes. 

\subsection{Experimental Set-up}
\label{sec:expsetup}

\paragraph*{Evaluation Measure} 
  
For our initial studies evaluating our design decisions and
  comparing our approach to three others (Yago, a simple tag-based
  baseline, and an Enterprise Knowledge Graph), we do not have a {\em
    user}.  We simulate a user by reporting the success
  probability of finding each table in the lake.   Conceptually what
  this means is that if a user {\em had in mind} a table that is in
  the lake and makes navigation decisions that favor picking states
  that are closest to attributes and tags of that table, we report the
  probability that they would find that table using our organization. 

We therefore report for our experiments a
  measure we call {\em success probability} that considers a navigation to
  be successful if it finds tables with an attribute of or a similar attribute to the query's attributes. 
  We first define the success probability for attributes. 
  Specifically, let $\kappa$ be a similarity measure between two attributes 
  and let $0 < \theta \leq 1$ be a similarity threshold. 
\begin{definition}
\label{def:nav}
The {\em success probability} of an attribute $\attA$ is
defined as
    \begin{equation}
Success(\attA|\mathcal{O}) = 1-\prod_{\attA_i\in \mathcal{A} \wedge
  \kappa(\attA_i, \attA) \geq \theta}  (1-P(\attA_i|\mathcal{O}))
    \end{equation}
\end{definition}
We use the Cosine similarity on the topic vectors of attributes for 
$\kappa$ and a threshold $\theta$ of $0.9$. 
Based on attribute success probabilities, we can compute table success probability as 
$Success(\attD|\mathcal{O}) = 1-\prod_{\attA\in \attD}(1-Success(\attA|\mathcal{O}))$. 
We report success probability for every table in the data
  lake sorted from lowest to highest probability on the x-axis (see
  Figure~\ref{fig:synthetic-orgs} as an example). 

\paragraph*{Implementation} Our implementation is in Python
and uses scikit-learn library for creating initial organizations.  Our
experiments were run on a 4-core Intel Xeon 2.6 GHz server with 128 GB
memory.  To speed up the evaluation of an organization, we cache the
similarity scores of attribute pairs as well as attribute and state pairs as states are updated during search. 

\subsection{Performance of Approximation}

We evaluate the effectiveness and efficiency of our exact algorithm
(using exact computation of discovery probabilities, not the
approximation discussed in Section~\ref{sec:approx}) 
on the {\tt TagCloud} benchmark. 

\subsubsection{Effectiveness}
\label{sec:synbenchmark-effectiveness}

\begin{figure}[t]
    \centering
\subfloat[\label{fig:synthetic-orgs}]{\includegraphics[width=0.49\columnwidth]{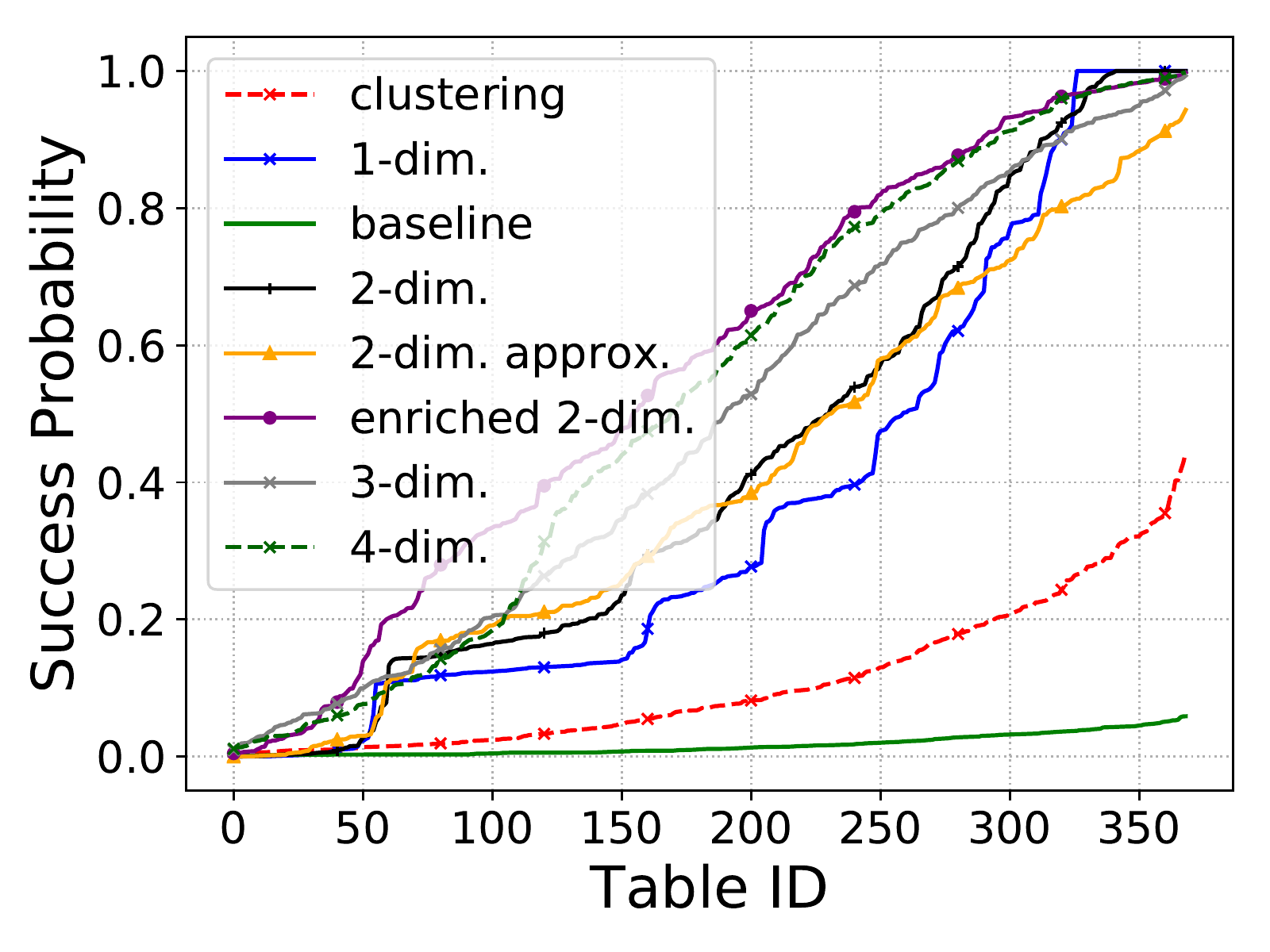}}
\centering
\subfloat[\label{fig:kb-orgs}]{\includegraphics[width=0.49\columnwidth]{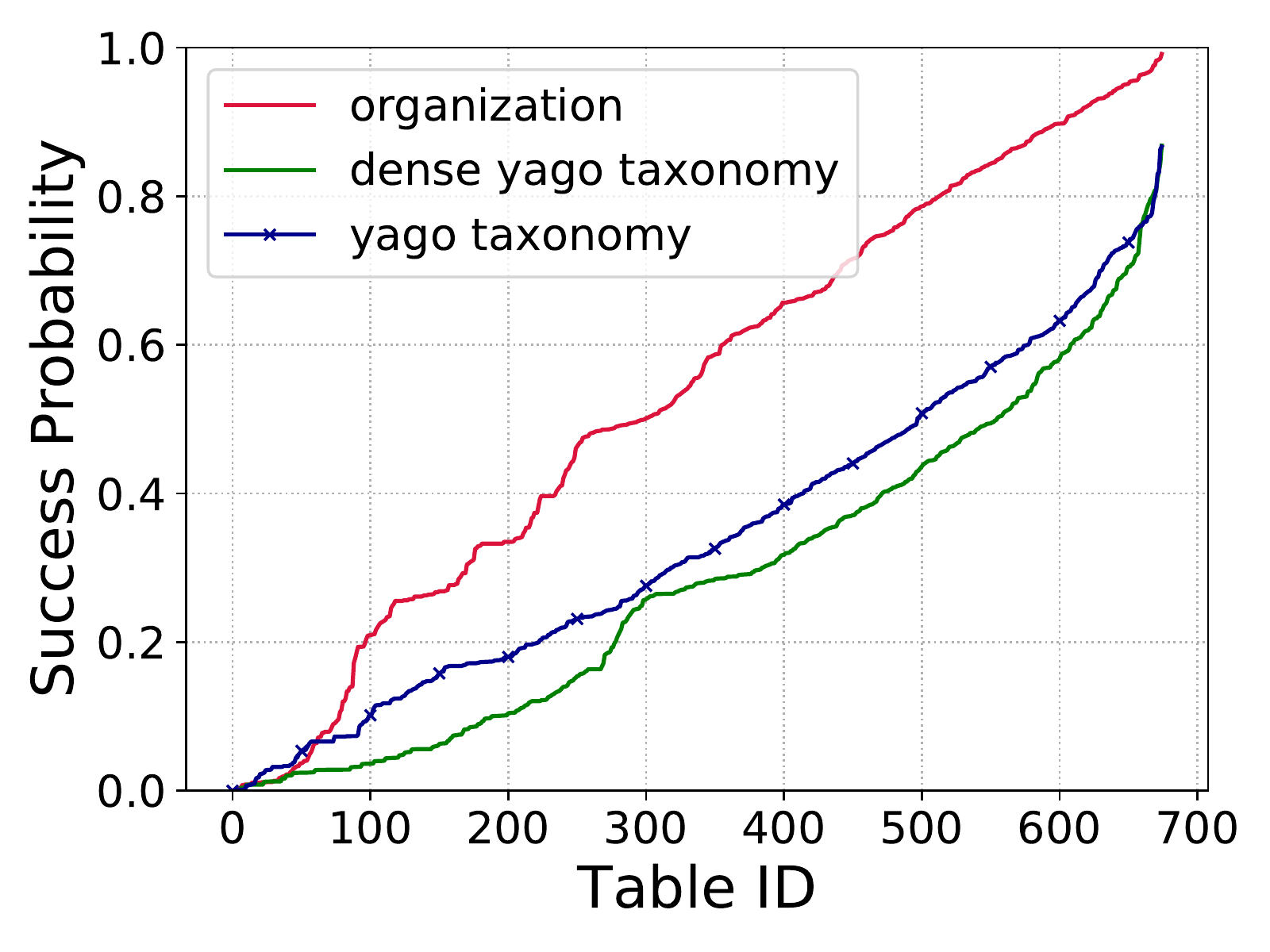}}
\caption{
	Success Probability of Organizations (a) on {\tt TagCloud} Benchmark and 
	(b) {\tt YagoCloud} Benchmark.}
\label{fig:synthetic-orgs}
\end{figure}

We constructed the {\tt baseline} organization where each
  attribute has as parents the tag-state of the attribute.  Recall in
  this benchmark each attribute has a single, accurate tag.  This
  organization is similar to the organization of open data portals.
We performed an agglomerative hierarchical clustering over this
baseline to create a hierarchy with branching factor 2.  This
organization is called {\tt clustering}. 
Then we used our algorithm to optimize the clustering organization to create  
N-dimensional ($N \in \{ 1, 2, 3, 4 \}$) organizations (called
  {\tt N-dim}). 
Figure~\ref{fig:synthetic-orgs} reports the success probability of each  
table in different organizations.  

In the {\tt baseline} organization, 
requires users to consider a large number of tags and select the best,
hence the average success probability for tables in this organization is just 0.016. 
This {\tt clustering} organization outperforms the {\tt baseline} by ten times.  
This is because the smaller branching factor of this organization reduces 
the burden of 
choosing among so many tags as the flat organization and results in 
larger transition probabilities to states even along lengthy paths. 
Our {\tt 1-dim} optimization of the {\tt clustering} organization
improves the success probability of the
{\tt clustering} organization by more than three times. 

To create the N-dimensional organization ($N > 1$), 
we clustered the tags into $N$ clusters (using $n$-medoids) and
  built an organization on each cluster.
The {\tt 2-dim} organzation 
has an average success probability
of 0.326 which is an improvement over the {\tt baseline} by 40 times. 
Although the number of initial tags is invariant 
among 1-dimensional and multi-dimensional organizations 
since each dimension is constructed on a smaller number of tags that 
are more similar, increasing the number of dimensions in an organization
improves the success probability, as shown in Figure~\ref{fig:synthetic-orgs}.  

In Figure~\ref{fig:synthetic-orgs}, almost 47 tables of {\tt TagCloud} have very 
low success probability in all organizations. 
We observed that almost 70\% of these tables contain only one attribute 
each of which is associated to only one tag. 
This makes these tables less likely to be discovered in any organization. 
To investigate this further, we augmented {\tt TagCloud} to
  associate each attribute with an additional tag 
  (the closest tag to the attribute other than its existing tag).  
We built a two-dimensional organization on the enriched {\tt TagCloud},
which we name {\tt enriched 2-dim}.   
This organization proves to have higher success probability 
overall, and improves the success probability for the least
  discoverable tables. 

\subsubsection{Efficiency}

\sloppypar{The construction time of {\tt clustering}, {\tt 1-dim}, {\tt 2-dim}, {\tt 3-dim}, 
{\tt 4-dim}, and {\tt enriched 2-dim} organizations 
are 0.2, 231.3, 148.9, 113.5, 112.7, 217 seconds, respectively.  
Note that the {\tt baseline} relies on the existing tags and 
requires no additional construction time. 
Since dimensions are optimized independently and in parallel, 
the reported construction times of the multi-dimensional organizations indicate 
the time it takes to finish optimizing all dimensions.}

\subsubsection{Approximation}

\begin{figure}[t]
    \centering
	\subfloat[\label{fig:domain-prunning}]{\includegraphics[width=0.49\columnwidth]{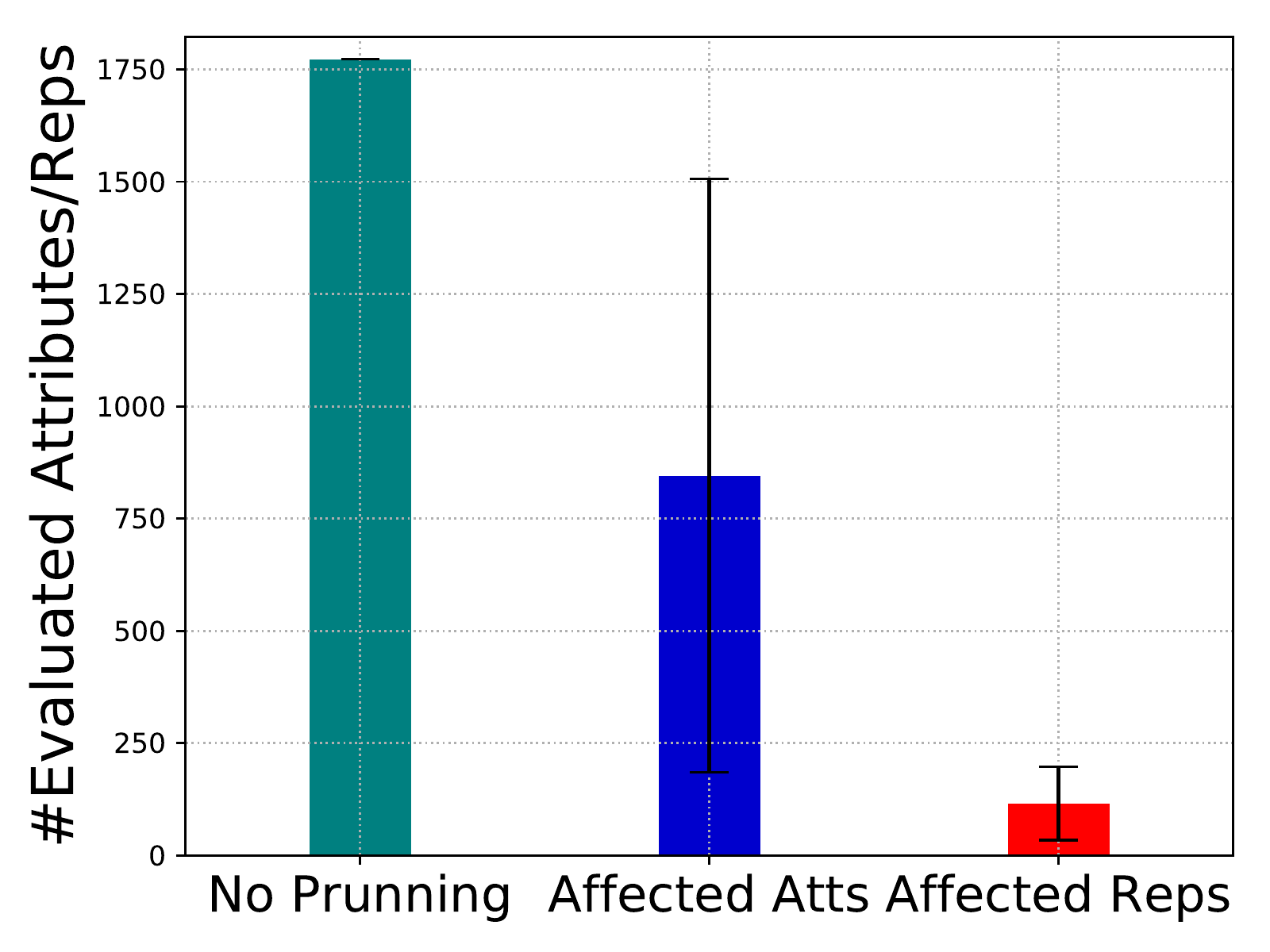}}
\centering
	\subfloat[\label{fig:state-prunning}]{\includegraphics[width=0.49\columnwidth]{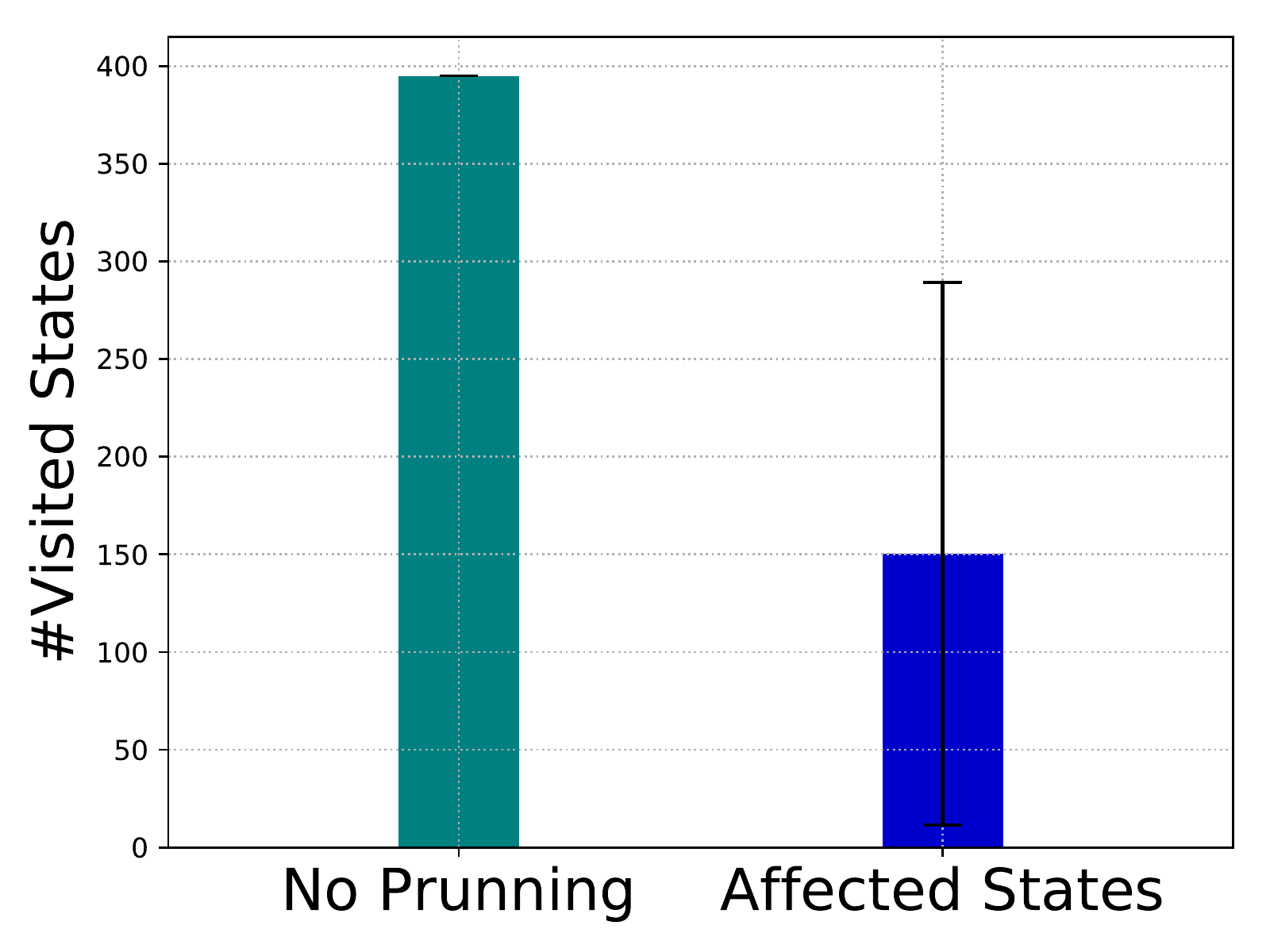}}
\caption{
	 Pruning (a) Domains and (b) States on {\tt TagCloud} Benchmark.}
\label{fig:synthetic-prunning}
\end{figure}

The effectiveness and efficiency numbers we have reported so far
  are   for the non-approximate version of our algorithm. 
To evaluate an organization during search 
we only examine the states and attributes that are affected by the change an operation has made. 
Thus, this pruning guarantees exact computation of success probabilities. 
Our experiments, shown in Figure~\ref{fig:synthetic-prunning},     
indicate that although local changes can potentially propagate 
to the whole organization,  
on average less than half of states and attributes are visited and evaluated for each
search iteration. 
Furthermore, we considered approximating discovery probabilities 
using a representative set size of  
10\% of the
attributes and only evaluated those representatives that correspond 
to the affected attributes. 
This reduces the number of discovery probability evaluations to only 6\% of the attributes. 
As shown in Figure~\ref{fig:synthetic-orgs}, named {\tt 2-dim approx},
this approximation has negligible impact on the success probabilities of tables in the constructed 
organization. 
The construction time of {\tt 2-dim} (without the approximation)
  is 148.9 seconds.  For {\tt 2-dim approx} (using a representative
  size of 10\%) is
  30.3 seconds.  The remaining experiments report organizations (on 
  much larger lakes) created using this approximation for
  scalability.

\subsection{Comparison of Organizations}
\label{sec:comparisons}

We now compare our approach with (1) a hand-curated taxonomy on the {\tt YagoCloud} benchmark,  
(2) the flat baseline on a large real data lake 
{\tt Socrata}, and 
(3) 
an automatically generated enterprise knowledge graph
(EKG)~\cite{fernandez2018aurum,fernandez2018seeping} on 
a smaller
sample of this lake  
{\tt Socrata-1}. 

\subsubsection{Comparison to A Knowledge Base Taxonomy}
\label{sec:kbtaxonomy}

We compare the 
effectiveness 
 of using the existing YAGO taxonomy on {\tt
  YagoCloud} tables for navigation with our organization.
Figure~\ref{fig:kb-orgs} shows the success probability of tables in the benchmark 
when navigating the YAGO taxonomy and our data lake organization. 
The taxonomy consists of 1,171 nodes and 2,274 edges 
while our 
(1-dimensional) 
organization consists of 442 nodes and 565 edges.  
Each state in the optimized organization consists of a set of types that need to be further interpreted  
while each state in the taxonomy refers to a YAGO class label. 
However, the more compact representation of topics of the attributes 
leads to a more effective organization for navigation. 
To understand the influence of taxonomy/organization size on
  discovery effectiveness, we condensed the taxonomy. 
Similar to the knowledge fragment selection approach~\cite{Hellmann:2008}, 
we created a compact taxonomy 
by eliminating the immediate level of nodes above the leaves unless they have other children.    
This leads to a taxonomy that has 444 nodes and 1,587 edges which is 
closer to the size of our 
organization. 
Condensing the taxonomy increases the branching factor of nodes while decreasing the length of discovery paths. 
This leads to a slight decrease in success probability compared to the original taxonomy.   
Knowledge base taxonomies are efficient for organizing the knowledge of entities. 
However, our organizations are able to optimize the navigation better
based on the data lake distribution.  

\subsubsection{Comparison to a Baseline on Real Data}

We constructed
ten organizations on the {\tt Socrata} lake by 
first partitioning its tags into ten groups using k-medoids clustering~\cite{kaufmann:1987}. 
We apply Algorithm~\ref{alg:organize} on each cluster to 
approximate an optimal organization. 
We use an agglomerative hierarchical clustering 
of  tags in {\tt Socrata} as the initial organization.
In each iteration, we approximate the success probability of the organization 
using a representative 
set with a size that is 10\%  of the total number of attributes in the organization. 
Table~\ref{tbl:orgstats} reports the number of representatives
considered for this approximation in each organization along with
other relevant statistics.
Since the cluster sizes are skewed, 
the number of attributes reachable via each organization has a high 
variance.  Recall that {\tt Socrata} has just over 50K attributes and they might have multiple tags, 
so many are reachable in multiple organizations.
It took 12 hours to construct the multi-dimensional organization. 

\begin{table}[h]
\small  
\centering  
\caption{Statistics of 10 Organizations of {\tt Socrata} Lake.} 
\label{tbl:orgstats}
\begin{tabular}{@{} l c c c c @{} }
\toprule  
\textbf{Org} & \textbf{\#Tags} & \textbf{\#Atts} & \textbf{\#Tables} & \textbf{\#Reps} \\
	\midrule  
	\textbf{1} & 2,031 & 28,248 & 3,284 & 2,824\\
	\textbf{2} & 1,735 & 11,363 & 1,885 & 1,136\\
	\textbf{3} & 1,648 & 20,172 & 9,792 & 2,017\\	
	\textbf{4} & 1,572 & 19,699 & 2,933 & 1,969\\
	\textbf{5} & 1,378 & 11,196 & 1,947 & 1,119\\
	\textbf{6} & 1,245 & 17,083 & 1,934 & 1,708\\	
	\textbf{7} & 829 & 8,848 & 1,302 & 884\\
	\textbf{8} & 353 & 6,816 & 831 & 681\\
	\textbf{9} & 240 & 3,834 & 614 & 383\\
	\textbf{10} & 43 & 118 & 33 & 11\\	
	\bottomrule  
\end{tabular}
\end{table} 

\begin{figure}[t]
    \centering
	\subfloat[\label{fig:od-orgs}]{\includegraphics[width=0.49\columnwidth]{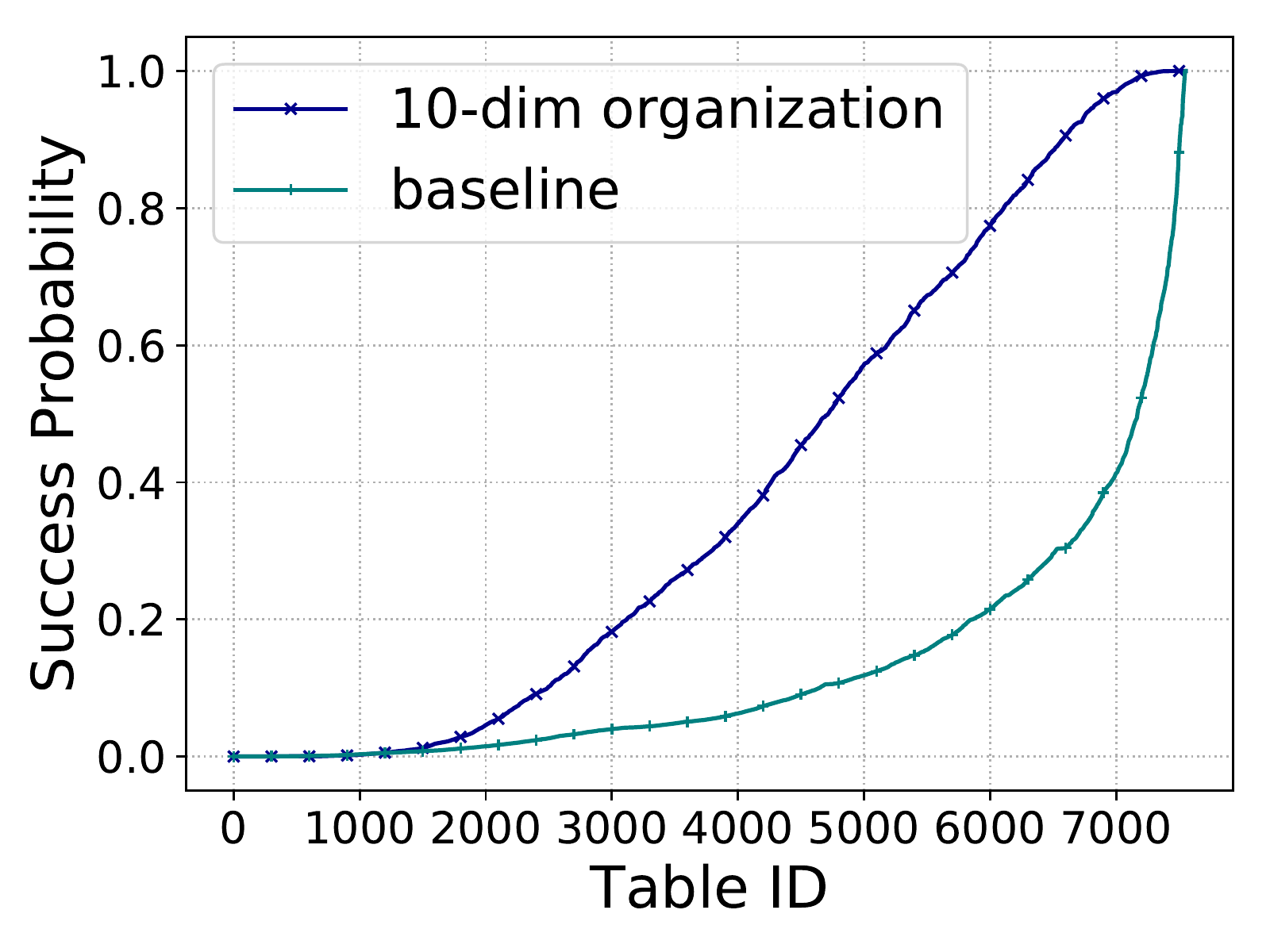}}
\centering
	\subfloat[\label{fig:ekg}]{\includegraphics[width=0.49\columnwidth]{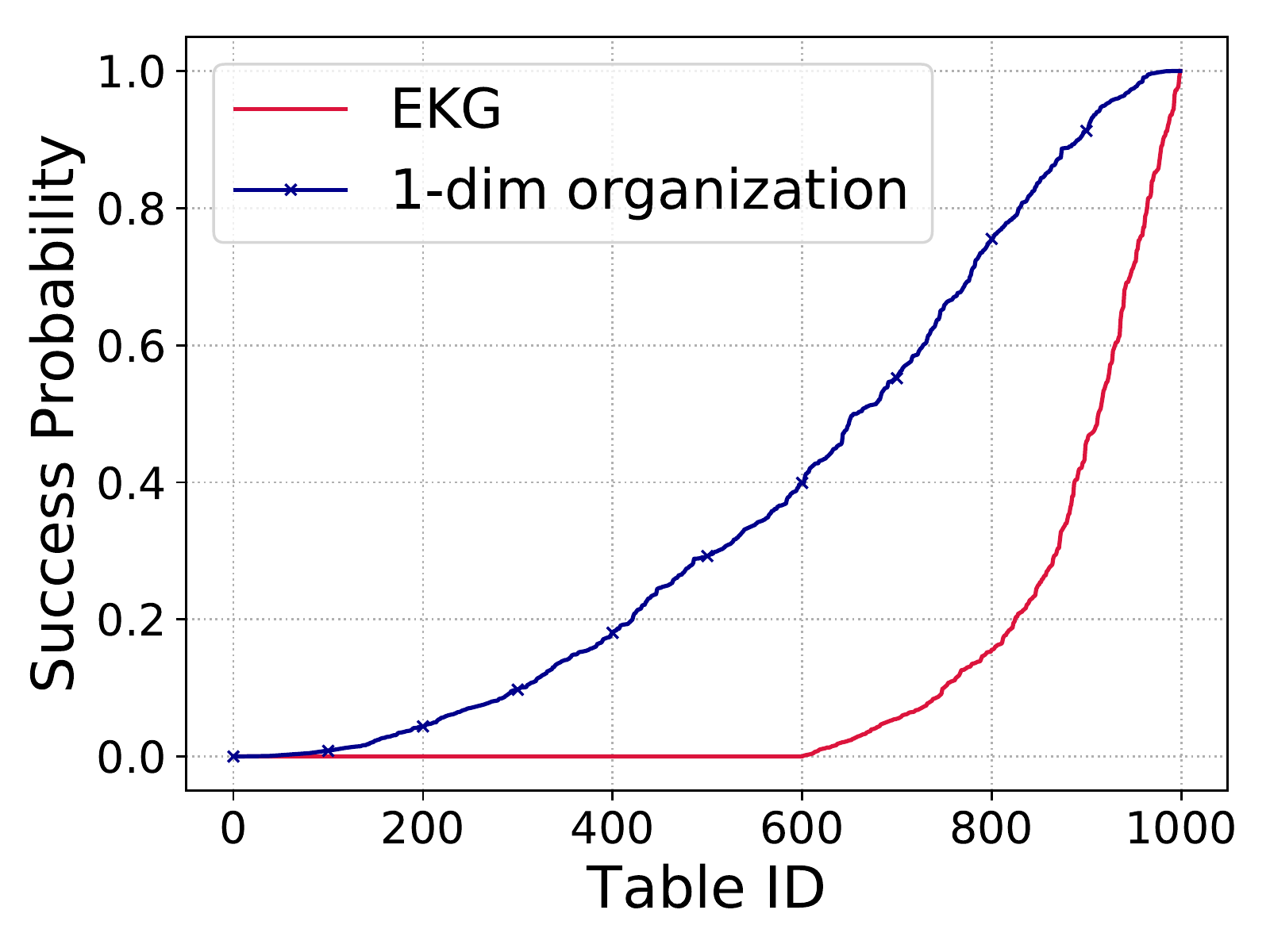}}
\caption{
	(a) Success Probability of Tables in {\tt Socrata} Lake using Organization and Baseline. (b)
        Success Probability of Tables in {\tt Socrata-1} Lake using
        Organization and EKG.}
\end{figure}

Figure~\ref{fig:od-orgs} shows the success probability of the ten organizations on the {\tt Socrata} data lake. 
Using this organization, a table is likely to be discovered during navigation of the data lake   
with probability of 0.38, compared to the current state of navigation in data portals using only tags,
which is 0.12. 
Recall that to evaluate the discovery probability of an attribute, we evaluate
the probability of discovering the  penultimate state that contains its tag 
 and  multiply it by the probability of selecting 
the attribute among the attributes associated to the tag. 
  The distribution of attributes to tags 
  depends on the metadata. 
Therefore, the organization algorithm does not have any control on the branching factor at the lowest level of the organization, 
  which means that the optimal organization 
is likely to have a lower success probability than 1. 

\subsubsection{Comparison to Linkage Navigation}
\label{sec:ekg}

Automatically generated linkage graphs are an alternative navigation structure to an organization. 
An example of a linkage graph is the Enterprise Knowledge Graph (EKG)~\cite{fernandez2018aurum,fernandez2018seeping}. 
Unlike organizations which facilitate exploratory navigation, 
in EKG, navigation starts with known data. 
A user writes queries on the graph to find related attributes through keyword search followed by similarity queries 
to find other attributes similar to these attributes.   

To study the differences between linkage navigation  
	and organization navigation, 
        we compare the probability of discovering tables using
        each.
We used a smaller lake ({\tt Socrata-1}) for this comparison 
because the abundance of linkages between attributes in a large data
lake makes our evaluation using large EKGs computationally expensive. 
In the EKG, each attribute in a data lake is a node of the graph
  and there are edges based on node similarity.  A syntactic edge
  means the Jaccard similarity of attribute
  values~\cite{fernandez2018aurum} is above a threshold and a semantic
  edge means the combined semantic and syntactic similarity of the
  attribute names is above a threshold~\cite{fernandez2018seeping}.  
To use EKG for navigation, we adapt Equation~\ref{eq:transprob} such   
	that the probability of a user navigating from node $m$ 
 to an adjacent node $s$ is proportional to the similarity of $s$ to $m$ and is penalized by the branching factor of $m$.  
We consider the similarity of $s$ and $m$ to be the maximum of their 
semantic similarity and syntactic similarity. 
Since the navigation can start from arbitrary nodes, 
	to compute the success probability of a table in an EKG, 
	we consider the average success probability of a table over up to 500  
	runs each starting from a random node. 
	We use the threshold $\theta=0.9$ for filtering the edges in EKG. 
	This makes 3,989 nodes reachable from some node in the graph. 
	The average and maximum branching factors of this EKG are 122.30 and 725, respectively. 
	The average success probability of 
is 0.0056. 
	
	Figure~\ref{fig:ekg} shows the success probability of 
	navigation using an EKG and using an organization built on {\tt Socrata-1}, 
	when we limited the start nodes to be the ancestors of attributes of a table.  
	Although the data lake organization has higher construction time (2.75 hours) 
	than
        the EKG (1.3 hours),  
	it outperforms EKG in effectiveness. 
	The EKG is designed for discovering similar attributes to a known attribute. 
	When EKG is used for exploration, the navigation can start from arbitrary nodes 
	which results in long discovery paths and low success probability. 
	Moreover, depending on the connectivity structure of an EKG, some attributes 
	are not reachable in any navigation run (points in the left side of Figure~\ref{fig:ekg}). 
	In an EKG the number of navigation choices at each node depends 
	on the distribution of attribute similarity. 
	This causes some nodes to have high branching factor and leads to overall low success probability.  

\subsection{Effectiveness of Metadata Enrichment}
\label{sec:metadataenrich}

\begin{figure}[t]
    \centering
	\subfloat[\label{fig:classsamples}]{\includegraphics[width=0.49\columnwidth]{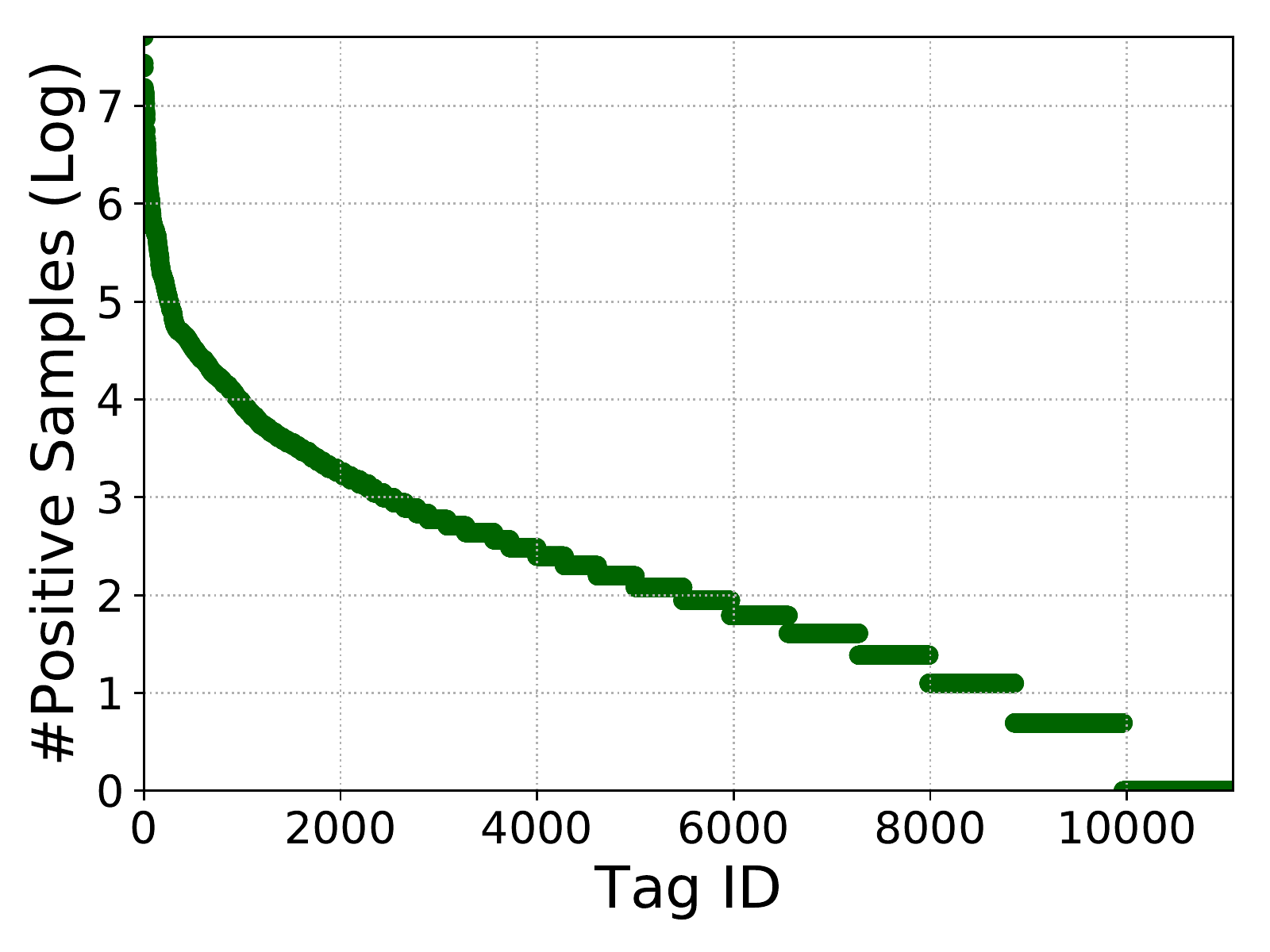}}
\centering
	\subfloat[\label{fig:od-classifiers}]{\includegraphics[width=0.49\columnwidth]{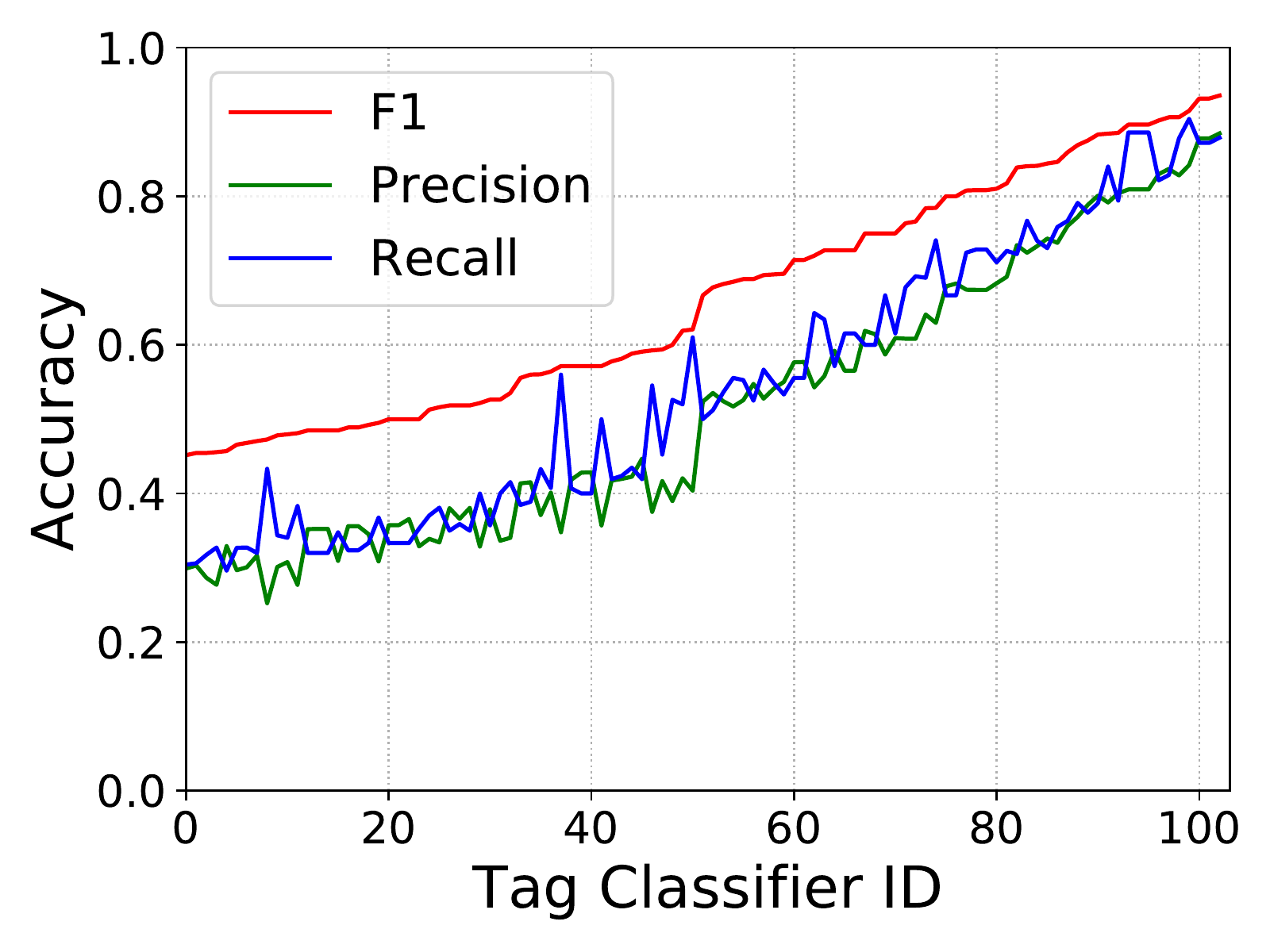}}
\caption{
	(a) No.~Positive Training Samples and \\ (b) Accuracy of Tag Classifiers.}
\label{fig:od-boosting}
\end{figure}

\begin{figure}[t]
    \centering
	\subfloat[\label{fig:tagdistckan}]{\includegraphics[width=0.49\columnwidth]{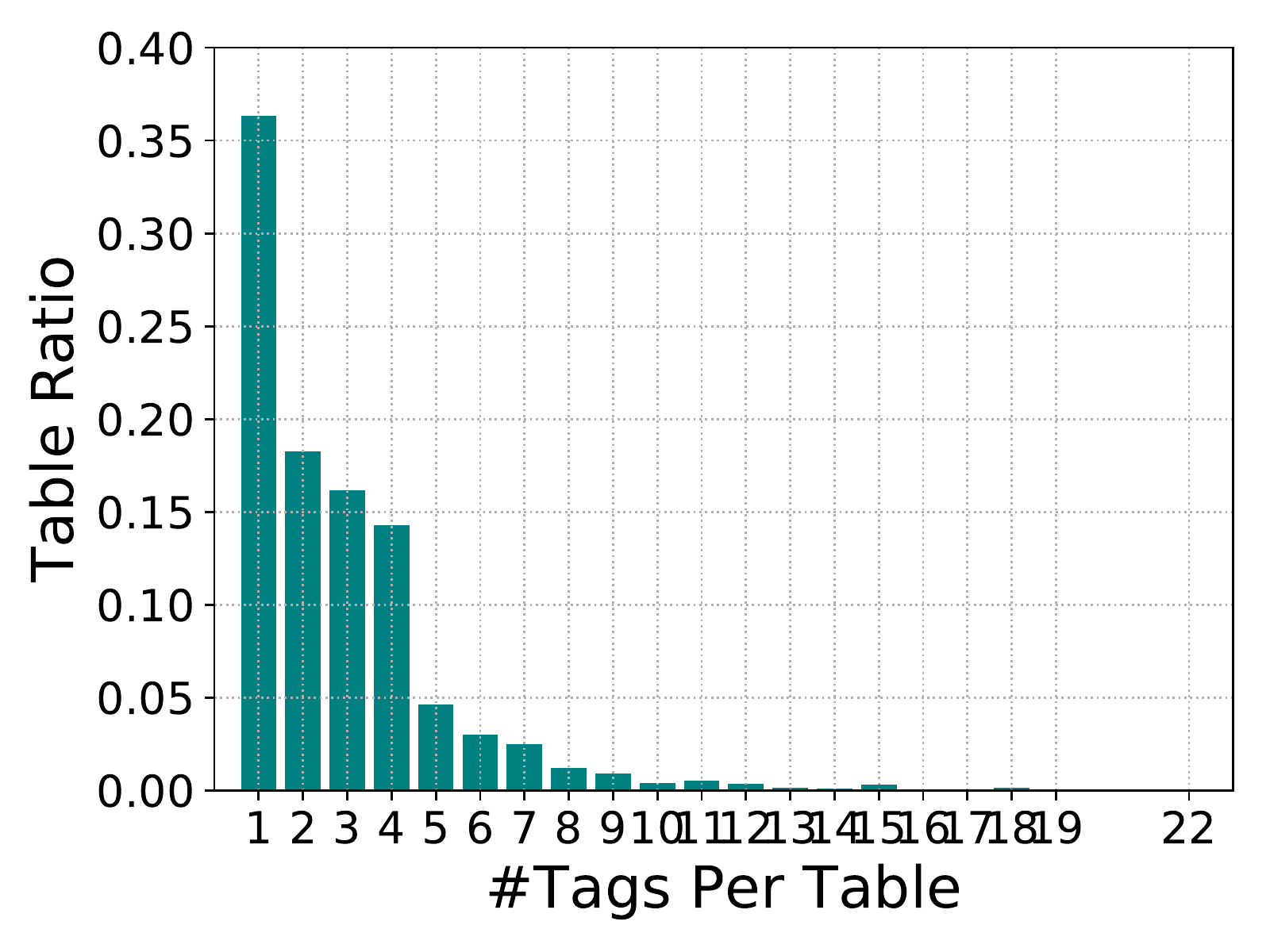}}
\centering
	\subfloat[\label{fig:ckanorg}]{\includegraphics[width=0.49\columnwidth]{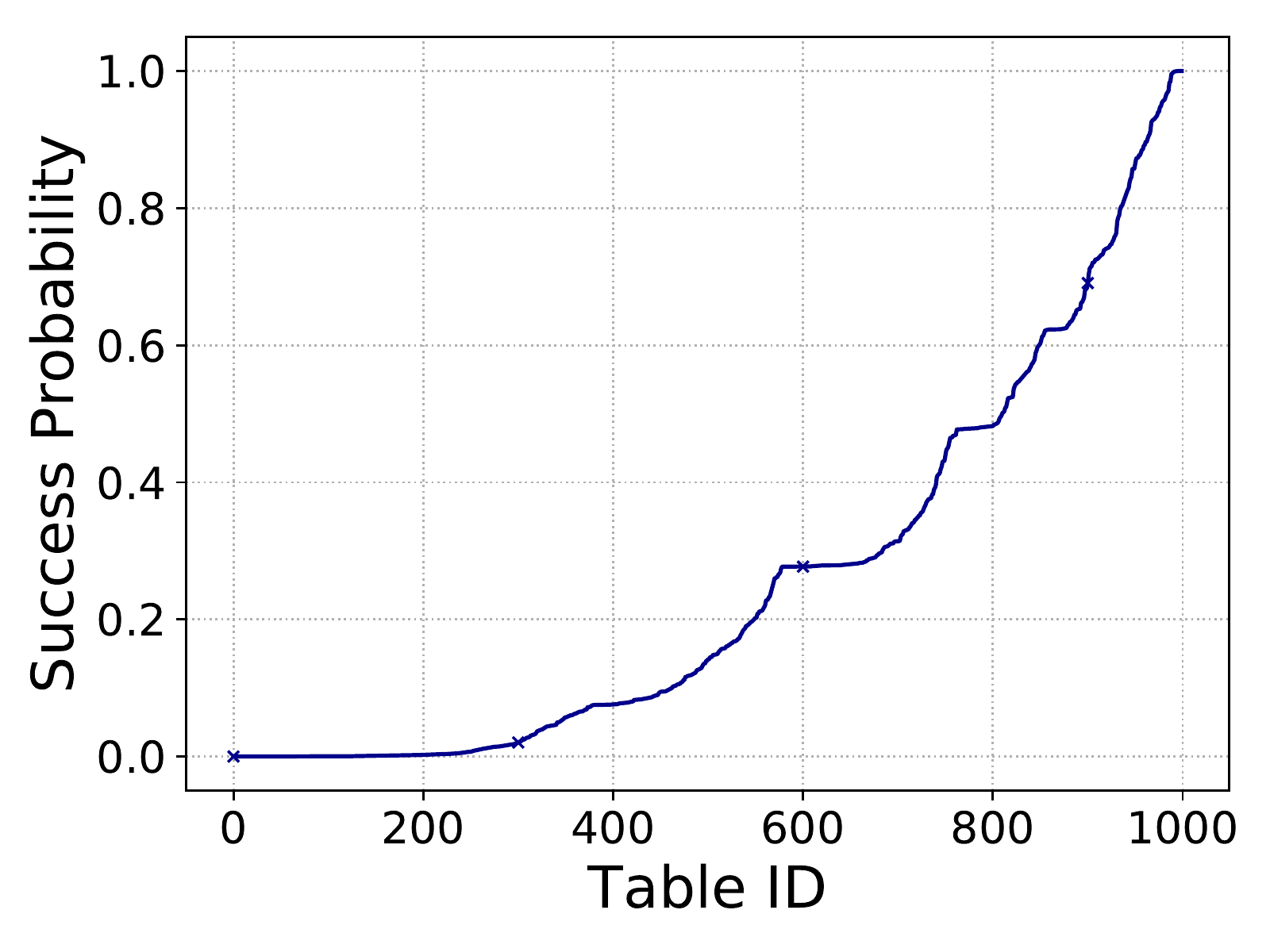}}
	\caption{
	(a) Distribution of Tags Added to {\tt CKAN}\\ 
 (b) Success Prob.~of {\tt CKAN} after Metadata Enrichment.}
\label{fig:od-boosting}
\end{figure}

The effectiveness of an organization depends on 
the number of tags associated to attributes. 
For data lakes with no or limited tags, we propose to transfer
tags from another data lake that has tags.   
Specifically, we transfer the tags of the {\tt Socrata} data lake to
attributes of {\tt CKAN} (which has no tags). 
Figure~\ref{fig:classsamples} shows the distribution of de-duplicated positive training samples per 
tag in {\tt Socrata}. 
Despite the large number of tags (11,083),
only very few of them have enough positive samples (associated attributes) 
for training a classifier. 
We considered the ratio of one to nine for positive and negative  samples. 
We employed the distributed gradient boosting of 
XGBoost~\cite{Chen:2016:XST} to train classifiers 
on the tags with at least 10 positive training samples (866 tags).
The training algorithm performs grid search for hyper-parameter tuning of classifiers.  
Figure~\ref{fig:od-classifiers} demonstrates the accuracy of the 10-fold cross validation of 
the classifiers with top-100 F1-scores. 
Out of 866 tags of {\tt Socrata}, 751 were associated to {\tt CKAN} attributes and a total of
7,347 attributes got at least one tag. 
The most popular tag is {\em domaincategory\_government}. 
Figure~\ref{fig:tagdistckan} shows the distribution of newly
associated tags to {\tt CKAN} attributes for the 20 most popular tags. 
Figure~\ref{fig:ckanorg} reports the success probability of a 1-dimensional organization. 
More than half of the tables are now searchable through the organization that were otherwise unreachable in the lake.  

\subsection{User Study}
\label{sec:userstudy}

We performed a formal user study to compare search by navigation 
to the major alternative of keyword search. 
Through a formal user study, we investigated how users perceive the usefulness of the two approaches. 

To remain faithful to keyword search engines, 
we created a semantic search engine that supports keyword search over attribute values and  
tables metadata (including attribute names and table tags). 
The search engine performs query expansion with semantically similar terms. 
The search
engine supports BM25~\cite{schutze2008introduction} document search and semantic keyword search using
pretrained GloVe word vectors \cite{pennington2014glove}.  
    Our implementation uses the library Xapian
    \footnote{https://xapian.org/} to perform keyword search.  Users can
    optionally enable query expansion by augmenting the keyword query with
    additional semantically similar terms. 
We also created a prototype that enables participants to navigate our organizations. 
In this prototype, each node of an organization is labeled with a set of representative tags. 
We label leaf nodes (where attributes are located) with corresponding table names 
and penultimate nodes (where single tags are located) with the corresponding tags. 
The remaining nodes are labeled 
with two tags which are the first and second most occurring tags, among its children's labels, 
for the nodes' attributes. 
If these tags belong to the label of the same child, we choose the third most occurring tag and so on. 
At each state, the user can 
navigate to a desired child  
node or backtrack to the parent of the current node. 

\textbf{Hypotheses.} The goal of our user study is to test the following hypotheses: H1) given the same amount of time,
participants would be able to find as many relevant tables with navigation as with keyword search; 
H2) given the same amount of 
time, participants who use navigation would be able to find relevant tables that cannot be found by keyword search.   
We measure the disjointness of results  
with the symmetric difference of result sets normalized by the 
size of the union of results, i.e., for two result sets $R$ and $T$, the disjointness is computed by
$1-\frac{|R \cap T|}{|R \cup T|}$. 

\textbf{Study Design.}
We considered {\tt Socrata-2} and {\tt Socrata-3} data lakes for our user study 
and defined an overview information need scenario for each data lake.  
We made sure that these scenarios are similar in difficulty 
	by asking a number of domain experts who were familiar with the underlying data lakes to 
	rate several candidate scenarios. 
Note that the statistically insignificant difference between the number of tabels found for each 
scenario by our participants provides evidence that our two scenarios were infact similar in difficulty.
The scenario used for {\tt Socrata-2} asks 
participants to find tables relevant to the scenario 
``suppose you are a journalist and you'd like to find datasets published by governments on the topic of \emph{smart city}.''. 
The scenario used for {\tt Socrata-3} asks participants to find tables relevant to the 
scenario ``Suppose you are a data scientist in a medical company and you would like to find 
datasets about \emph{clinical research}''.  
During the study, we 
asked participants to use keyword search and navigation to find a set of tables which they deemed relevant to given 
scenarios. 
For this study, we recruited 12 participants using convenience sampling~\cite{EtikanMA2016,given2008sage}. The participants have diverse backgrounds with undergraduate and graduate education in 
computer science, engineering, math, statistics, and management sciences.  

This study was a {\em within-subject} design. 
In our setting, we want to avoid 
two potential sources of invalidity. 
First, participants might become familiar with the underlying
data lake during the first scenario which then might 
help them to search better in their second scenario. 
To address this problem, we made sure that {\tt Socrata-2} and {\tt Socrata-3} 
do not have overlapping tags and tables.    
Second, the sequence in which the participants
use the two search approaches can be the source of confounding. To mitigate for this problem,
we made sure that half of the participants first performed keyword search and the rest performed  
navigation first. In summary, we handled the effect of these two sources of invalidity using
a balanced \emph{latin square} design with 4 blocks (block 1: {\tt Socrata-2}/navigation first, block 2: 
{\tt Socrata-3}/navigation first, etc.). 
We randomly assigned equal number of participants to each of these blocks. For each
participant, the study starts with a short training session, after
that, we gave the participants 20 minutes for each scenario. 
\begin{table}
\small
\centering
	\caption{Number of Tables Found by the Search Techniques.}
	\label{tbl:userstudydata1}
\begin{tabular}{@{} l c c @{} }
\toprule  
\textbf{Participant ID} & \textbf{Navigation} & \textbf{Keyword Search}\\
	\midrule  
	\textbf{1} & 11 & 8\\
	\textbf{2} & 32 & 24\\
	\textbf{3} & 2 & 2\\	
	\textbf{4} & 25 & 13\\
	\textbf{5} & 45 & 14\\
	\textbf{6} & 44 & 34\\	
	\textbf{7} & 23 & 10\\
	\textbf{8} & 14 & 8\\
	\textbf{9} & 20 & 10\\
	\textbf{10} & 7 & 8\\
	\textbf{11} & 6 & 1\\	
	\textbf{12} & 6 & 2\\	
	\bottomrule  
	\end{tabular}
\end{table} 

\textbf{Results.}
Because of our small sample size, we used the
non-parametric Mann-Whitney test to determine the
significance of the results and tested our two-tailed
hypotheses. We found that there is no statistically significant
difference between the number of relevant tables found using the organization and keyword search. 
The number of tables found by each participant for each scenario is listed in Table~\ref{tbl:userstudydata1}. 
We first asked two collaborators to eliminate irrelevant tables found. Since the
number of irrelevant data was negligible (less than 1\% for both approaches) 
we will not further report on this process. 
This confirms our first hypothesis. 
The maximum number of tables found  by navigating an organization and performing keyword search was 44 and 34, respectively. 
Moreover, a Mann-Whitney test indicated that the disjointness of 
results was greater for participants who used organization (Mdn = 0.985) than for 
participants who used keyword search (Mdn = 0.916), U = 612, p = 0.0019. 
This confirms our second hypothesis. Note that the disjointness was computed for each pair of participants who
 worked on the same scenario using the same approach, then the pairs generated for each technique were compared together. 
 The calculated disjointness pairs are included in Table~\ref{tbl:userstudydata2}. 
Based on our investigation, 
this difference might be because participants used very similar keywords, 
whereas the paths which were
taken by each participant while navigating an organization were very different. 
In other words, 
As some participants described, they were having a hard time finding keywords 
that best described their interest since 
they did not know what was available, 
whereas with our organization, at each step, 
they could see what seemed more interesting to them and find their way based on their preferences.
As one example, for the {\em smart city}  overview scenario, everyone found
tables tagged with the term {\tt City} using search.  But using
organization, some users found traffic monitoring data, while others
found crime detection data, while others found renewable energy
plans.  
Of course if they knew {\em a priori} this data was in the
lake they could have formulated better keyword search queries, but
navigation allowed them to conveniently discover these relevant
tables without prior knowledge.
One very interesting observation in this study is that 
although participants find similar number of tables, 
there is only around 5\% intersection between tables found using
keyword search and tables found using our approach. 
This suggests that organization can be a good complement to the keyword search and 
vice versa.

We evaluated the usability by asking each participant
to fill out a standard post-experiment system usability
  scale (SUS) questionnaire~\cite{Brooke:2013:SR:2817912.2817913} after each block.  
This questionnaire is designed to measure a user's judgment of a
  system's effectiveness and efficiency. 
 We analyzed participants' rankings, and kept record of the number of questions 
 for which they gave a higher rankings to each approach.
 Our results indicate that 58\% of the participants preferred to use keyword search,
we suspect in part due to familiarity.  No participant had neutral preference. Still having 42\% prefer
  navigation indicates a clear role for this second, complementary modality. 

\begin{table*}
	\centering
\caption{Disjointness Scores of Result Sets of Keyword Search (kw) and Navigation (nav) for each Pair of Participants.}
\label{tbl:userstudydata2}
\resizebox{\textwidth}{!}{
	\begin{tabular}{|>{\columncolor[HTML]{FFFFFF}}c |l|l|l|l|l|l|l|l|l|l|l|l|}
\hline
	\textbf{Participant ID} & \multicolumn{1}{c|}{\cellcolor[HTML]{FFFFFF}\textbf{\begin{tabular}[c]{@{}c@{}}1\\ Kw/Nav\end{tabular}}} & \multicolumn{1}{c|}{\cellcolor[HTML]{FFFFFF}\textbf{\begin{tabular}[c]{@{}c@{}}2\\ Kw/Nav\end{tabular}}} & \multicolumn{1}{c|}{\cellcolor[HTML]{FFFFFF}\textbf{\begin{tabular}[c]{@{}c@{}}3\\ Kw/Nav\end{tabular}}} & \multicolumn{1}{c|}{\cellcolor[HTML]{FFFFFF}\textbf{\begin{tabular}[c]{@{}c@{}}4\\ Kw/Nav\end{tabular}}} & \multicolumn{1}{c|}{\cellcolor[HTML]{FFFFFF}\textbf{\begin{tabular}[c]{@{}c@{}}5\\ Kw/Nav\end{tabular}}} & \multicolumn{1}{c|}{\cellcolor[HTML]{FFFFFF}\textbf{\begin{tabular}[c]{@{}c@{}}6\\ Kw/Nav\end{tabular}}} & \multicolumn{1}{c|}{\cellcolor[HTML]{FFFFFF}\textbf{\begin{tabular}[c]{@{}c@{}}7\\ Kw/Nav\end{tabular}}} & \multicolumn{1}{c|}{\cellcolor[HTML]{FFFFFF}\textbf{\begin{tabular}[c]{@{}c@{}}8\\ Kw/Nav\end{tabular}}} & \multicolumn{1}{c|}{\cellcolor[HTML]{FFFFFF}\textbf{\begin{tabular}[c]{@{}c@{}}9\\ Kw/Nav\end{tabular}}} & \multicolumn{1}{c|}{\cellcolor[HTML]{FFFFFF}\textbf{\begin{tabular}[c]{@{}c@{}}10\\ Kw/Nav\end{tabular}}} & \multicolumn{1}{c|}{\cellcolor[HTML]{FFFFFF}\textbf{\begin{tabular}[c]{@{}c@{}}11\\ Kw/Nav\end{tabular}}} & \multicolumn{1}{c|}{\cellcolor[HTML]{FFFFFF}\textbf{\begin{tabular}[c]{@{}c@{}}12\\ Kw/Nav\end{tabular}}} \\ \hline

\textbf{1} &  &  & 0.733/1.000 & 0.714/0.937 & 0.916/1.000 & 0.875/1.000 &  &  &  &  & 0.947/1.000 & 0.800/0.928 \\ \hline
\textbf{2} &  &  &  &  &  &  & 1.000/1.000 & 1.000/1.000 & 1.000/1.000 & 1.000/1.000 &  &  \\ \hline
\textbf{3} & 0.733/1.000 &  &  & 0.850/0.960 & 0.941/1.000 & 0.923/0.966 &  &  &  &  & 0.850/0.985 & 0.850/1.000 \\ \hline
\textbf{4} & 0.714/0.937 &  & 0.850/0.960 &  & 1.000/1.000 & 0.818/1.000 &  &  &  &  & 0.871/0.985 & 0.777/0.837 \\ \hline
\textbf{5} & 0.916/1.000 &  & 0.941/1.000 & 1.000/1.000 &  & 1.000/0.833 &  &  &  &  & 1.000/1.000 & 1.000/1.000 \\ \hline
\textbf{6} & 0.875/1.000 &  & 0.923/0.966 & 0.818/1.000 & 1.000/0.833 &  &  &  &  &  & 0.939/0.979 & 0.916/0.964 \\ \hline
\textbf{7} &  & 1.000/1.000 &  &  &  &  &  & 0.913/0.969 & 0.892/0.930 & 0.935/1.000 &  &  \\ \hline
\textbf{8} &  & 1.000/1.000 &  &  &  &  & 0.913/0.969 &  & 0.888/1.000 & 0.909/1.000 &  &  \\ \hline
\textbf{9} &  & 1.000/1.000 &  &  &  &  & 0.892/0.930 & 0.888/1.000 & & 0.875/0.969 &  &  \\ \hline
\textbf{10} &  & 1.000/1.000 &  &  &  &  & 0.935/1.000 & 0.909/1.000 & 0.875/0.969 &  &  &  \\ \hline
\textbf{11} & 0.947/1.000 &  & 0.850/0.985 & 0.871/0.985 & 1.000/1.000 & 0.939/0.979 &  &  &  &  &  & 0.952/0.901 \\ \hline
\textbf{12} & 0.800/0.928 &  & 0.850/1.000 & 0.777/0.837 & 1.000/1.000 & 0.916/0.964 &  &  &  &  & 0.952/0.901 &  \\ \hline
\end{tabular}
	}
\end{table*}

\section{Related Work}
\label{sec:relatedwork}

 {\bf Entity-based Querying -} While traditional search engines are built for pages and keywords, 
 in entity search, a user formulates queries 
 to directly describe her target entities and 
 the result is a collection of 
 entities that match the query~\cite{Chakrabarti:2006}. 
 An example of a query is ``database \#{\tt professor}'', 
 where {\tt professor} is the target entity type and ``database'' is 
 a descriptive keyword.
 Cheng et. al propose a ranking algorithm for the result of entity queries 
 where a user's query is described 
 by keywords that may appear in the context of desired entities~\cite{Cheng:2007}. 

{\bf Data Repository Organization -} Goods is Google's specialized dataset catalog 
for about billions of Google's internal datasets~\cite{Halevy:2016}. 
The main focus of Goods is to collect and infer metadata for a large repository of datasets and 
make it searchable using keywords. 
Similarly, IBM's LabBook provides rich collaborative metadata 
graphs on enterprise data lakes~\cite{Labbook15}. 
Skluma~\cite{BSCF17} also extracts metadata graphs from a file system of datasets. 
Many of these {\em metadata} approaches include the use of static or 
dynamic linkage graphs~\cite{Deng+17,Man+18,fernandez2018seeping,fernandez2018aurum}, 
join graphs for adhoc navigation~\cite{DBLP:journals/pvldb/ZhuPNM17}, 
or version graphs~\cite{HellersteinSGSA17}. 
These graphs allow navigation from dataset to dataset.
However, none of these approaches learn new navigation structures optimized for dataset discovery.

{\bf Taxonomy Induction -} 
The task of taxonomy induction creates hierarchies where edges
represent {\em is-a} (or subclass) relations between classes.
The {\em is-a} relation represents true abstraction, not just the {\em subset-of} relation 
as in our approach. 
Moreover, taxonomic relationship between two classes exists independent of 
the size and distribution of the data being organized.  
As a result, taxonomy induction relies on 
ontologies or semantics
 extracted from text~\cite{Kozareva:2010} or structured data~\cite{PoundPT11}.
Our work is closes to concept learning, 
where entities are grouped into new concepts that are themselves organized in {\em is-a} 
hierarchies~\cite{Lehmann:2009:DLC:1577069.1755874}.

{\bf Faceted Search -} 
Faceted search enables the exploration of entities 
by 
refining the search results based on some properties or facets.  
A facet consists of a predicate (e.g., {\tt model}) and a set of possible terms 
(e.g., {\tt honda, volvo}). 
The 
facets may or may not have a hierarchical relationship. 
Currently, most successful faceted search systems rely on 
term hierarchies that are either designed manually by domain experts 
or automatically created using methods similar to 
taxonomy induction~\cite{zheng2013survey,Pound:2011:FDS,Duan:2013:SKS}. 
The large size and dynamic nature of data lakes makes 
the manual creation of a hierarchy 
infeasible. 
Moreover, since values in tables 
may not exist in 
external corpora~\cite{NargesianZPM18}, 
such taxonomy construction approaches 
of limited usefulness for 
the data lake organization 
problem.  

{\bf Keyword Search - } Google's dataset search uses keyword search
over metadata and relies on dataset owners providing rich semantic
metadata~\cite{DBLP:conf/www/BrickleyBN19}.  As shown in our user
study this can help users who know what they are looking for, but has
less value in serendipitous data discovery as a user tries to better
understand what data is available in a lake.

\section{Conclusion and Future Work}
\label{sec:conclusion}

We defined the data lake organization problem 
of creating an optimal organization over tables in a data lake.
We proposed a probabilistic framework that models
navigation in data lakes on an organization graph.  We frame the
data lake organization problem as an optimization problem of finding 
an organization  that maximizes the discovery probability of tables in a data
lake and proposed an efficient approximation
algorithm for creating good organizations. 
To build an organization, we use the attributes of tables  
together with any tags over the tables to combine table-level
  and instance-level features. 
The effectiveness and efficiency of our system are evaluated by
benchmark experiments involving synthetic and real world datasets. We
have also conducted a user study where participants use our system and a keyword search
engine to perform the same set of tasks.  It is shown that our system
offers good performance, and complements keyword search engines in data exploration. 
Future work includes empirically studying the scalability of our algorithm to larger data lakes, 
integrating keyword search and navigation as two interchangable modalities in a unified data
exploration framework. Based on the feedback and comments from the user study, we strongly believe
that these extensions will further improve the user's ability to navigate in
large data lakes.

\clearpage
\bibliographystyle{spmpsci}  
\bibliography{main}
\end{document}